\documentclass{vgtc}                          %

\onlineid{1397}

\title{TouchWalker: Real-Time Avatar Locomotion\\from Touchscreen Finger Walking}

\usepackage{relsize}

\author{
  Geuntae Park\thanks{e-mail: qkrrmsxo01@hanyang.ac.kr}\\
  {\smaller[1.0] Hanyang University}
  \and
  Jiwon Yi\thanks{e-mail: babap8514@hanyang.ac.kr}\\
  {\smaller[1.0] Hanyang University}
  \and
  Taehyun Rhee\thanks{e-mail: taehyun.rhee@unimelb.edu.au}\\
  {\smaller[1.0] University of Melbourne}
  \and
  Kwanguk Kim\thanks{e-mail: kenny@hanyang.ac.kr}\\
  {\smaller[1.0] Hanyang University}
  \and
  Yoonsang Lee\thanks{e-mail: yoonsanglee@hanyang.ac.kr (corresponding author)}\\
  {\smaller[1.0] Hanyang University}
}

\abstract{%
  
\textcolor{rv}{We present TouchWalker, a real-time system for controlling full-body avatar locomotion using finger-walking gestures on a touchscreen. The system comprises two main components: TouchWalker-MotionNet, a neural motion generator that synthesizes full-body avatar motion on a per-frame basis from temporally sparse two-finger input, and TouchWalker-UI, a compact touch interface that interprets user touch input to avatar-relative foot positions. Unlike prior systems that rely on symbolic gesture triggers or predefined motion sequences, TouchWalker uses its neural component to generate continuous, context-aware full-body motion on a per-frame basis—including airborne phases such as running, even without input during mid-air steps—enabling more expressive and immediate interaction. To ensure accurate alignment between finger contacts and avatar motion, it employs a MoE-GRU architecture with a dedicated foot-alignment loss. We evaluate TouchWalker in a user study comparing it to a virtual joystick baseline with predefined motion across diverse locomotion tasks. Results show that TouchWalker improves users’ sense of embodiment, enjoyment, and immersion.
}

}

\teaser{
  \centering
  \includegraphics[trim=125 13 125 50, clip, width=0.11\linewidth]{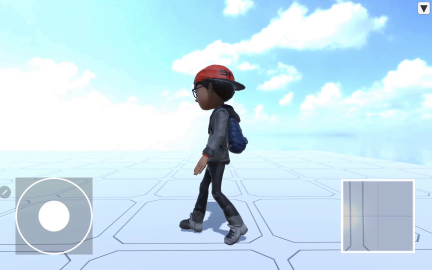}
  \includegraphics[trim=125 13 125 50, clip, width=0.11\linewidth]{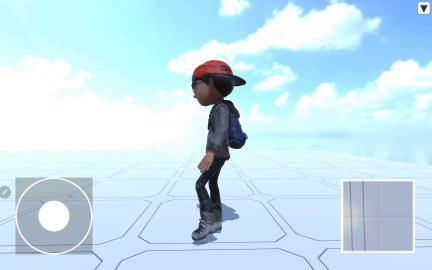}
  \includegraphics[trim=125 13 125 50, clip, width=0.11\linewidth]{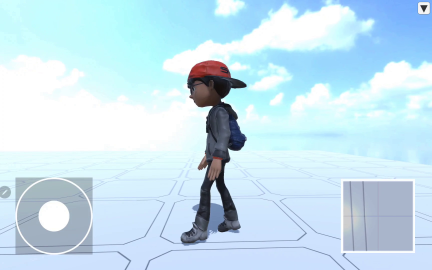} 
  \includegraphics[trim=125 13 125 50, clip, width=0.11\linewidth]{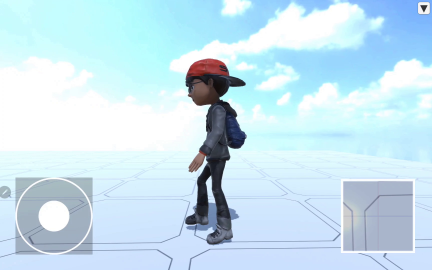}
  \includegraphics[trim=125 13 125 50, clip, width=0.11\linewidth]{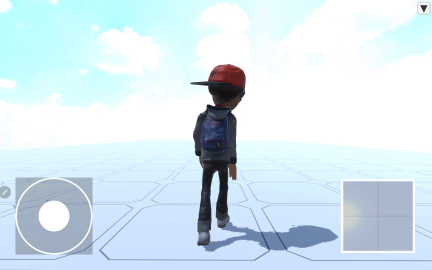}
  \includegraphics[trim=125 13 125 50, clip, width=0.11\linewidth]{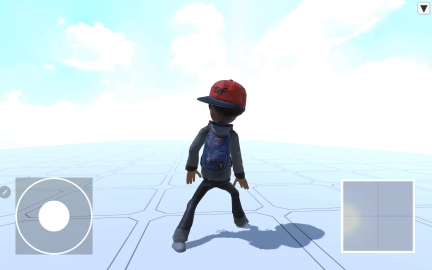}
  \includegraphics[trim=125 13 125 50, clip, width=0.11\linewidth]{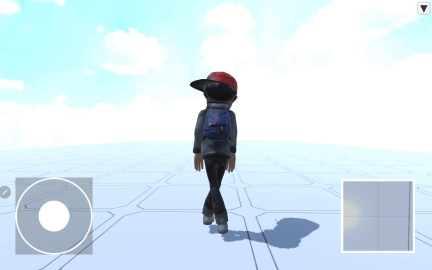} 
  \includegraphics[trim=125 13 125 50, clip, width=0.11\linewidth]{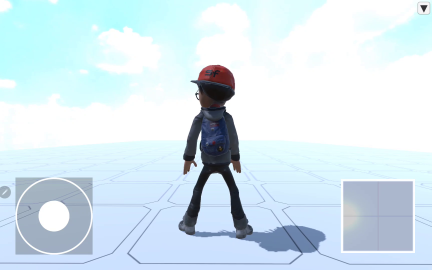}
  \includegraphics[trim=100 38 125 63, clip, width=0.11\linewidth]{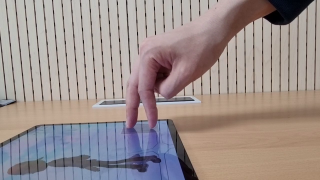}
  \includegraphics[trim=100 38 125 63, clip, width=0.11\linewidth]{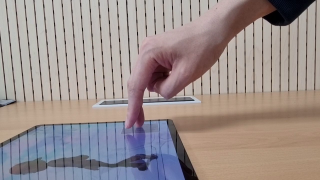}
  \includegraphics[trim=100 38 125 63, clip, width=0.11\linewidth]{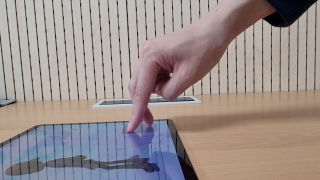}
  \includegraphics[trim=100 38 125 63, clip, width=0.11\linewidth]{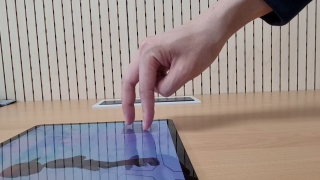}
  \includegraphics[trim=100 18 88 52, clip, width=0.11\linewidth]{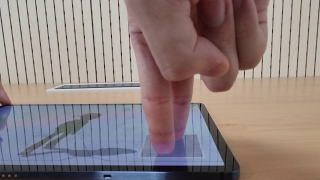}
  \includegraphics[trim=100 18 88 52, clip, width=0.11\linewidth]{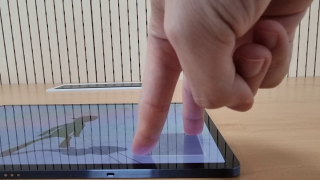}
  \includegraphics[trim=100 18 88 52, clip, width=0.11\linewidth]{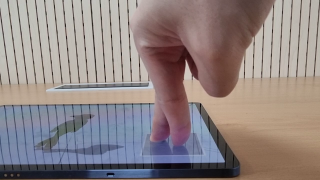}
  \includegraphics[trim=100 18 88 52, clip, width=0.11\linewidth]{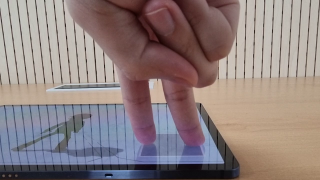}
  \caption{TouchWalker enables real-time control of full-body avatar locomotion using finger walking on a touchscreen. Users specify foot contacts within a touch region overlaid on the screen (bottom), and the system generates responsive full-body locomotion (top).
  }
  \label{fig:teaser}
}

\graphicspath{{figs/}{figures/}{pictures/}{images/}{./}} %

\usepackage{tabu}                      %
\usepackage{booktabs}                  %
\usepackage{lipsum}                    %
\usepackage{mwe}                       %
\usepackage{kotex}
\usepackage{amsfonts}
\usepackage{amsmath}
\usepackage{siunitx}
\usepackage{subcaption}

\usepackage{mathptmx}                  %

\usepackage{xcolor} 
\definecolor{rv}{rgb}{0,0,0}     %
\definecolor{rv2}{rgb}{0,0,0}     %

\begin{document}

\firstsection{Introduction}
\maketitle

In many forms of digital interactive content, users interact with on-screen avatars, making effective avatar control interfaces a fundamental component of the user experience.
On touchscreen devices, 
a virtual joystick is a commonly used to simulate an analog stick on the screen 
\cite{baxter_virtual_2023,teather_tilt-touch_2017,zaman_touchscreens_2010}.
While widely adopted, this interface primarily mimics traditional input devices and may not fully leverage the expressive potential of touchscreen interaction, such as variable input locations and multi-finger gestures.
Moreover, the lack of tactile feedback in virtual joysticks may limit users’ sense of embodiment and immersion, particularly in fine-grained or continuous movement control,
as tactile cues have been shown to enhance immersive experience and strengthen the connection between users and virtual avatars 
\cite{sun_augmented_2022,li_haptics-mediated_2024}.

{\color{rv}
To address these limitations, we explore finger walking as a more embodied and tactile interaction method for touchscreens. It leverages rhythmic bi-finger gestures to evoke the sensation of foot-ground contact, offering rich temporal and spatial cues for avatar control. However, prior systems have not fully exploited this potential—typically interpreting gestures symbolically or mapping them to predefined motion sequences—thus offering limited support for responsive, continuous full-body motion generation.

In this paper, we propose TouchWalker, a real-time system that enables expressive full-body avatar control through finger-walking gestures on a touchscreen. The system comprises two main components. At its core is TouchWalker-MotionNet, a neural motion generator that synthesizes full-body locomotion on a per-frame basis from temporally sparse two-finger input. Built on this foundation, TouchWalker-UI is a compact interface that interprets user touch input as avatar-relative foot positions, enabling responsive and embodied control. At each frame, the system receives the horizontal position of the stance foot—specified by the user’s finger—and generates full-body motion using TouchWalker-MotionNet, trained on motion capture data. To support finger-walking gestures within a limited touch area, the input is interpreted relative to the avatar’s coordinate space rather than absolute screen position.
This design enables finger walking to serve not just as a gesture trigger, but as a continuous and expressive control modality—marking a practical shift in how such input is leveraged for embodied avatar control.}

Real-time control using finger-walking input, however, poses several challenges. The system receives only the position of the stance foot, requiring it to infer the swing foot motion and full-body pose without explicit input. Furthermore, dynamic actions such as running involve airborne phases where no contact occurs, making it necessary to generate plausible motion even in the absence of touch input.
To address these challenges, we design \textcolor{rv}{TouchWalker-MotionNet,} a neural network that combines a Mixture of Experts (MoE) architecture with Gated Recurrent Units (GRU), enabling the model to incorporate temporal context and generate coherent full-body motion from sparse input. We also introduce a foot-alignment loss that encourages consistency between the synthesized motion and the user’s finger input.

\textcolor{rv}{
We conducted a user study comparing TouchWalker—which generates avatar motion in real time from finger input—with a virtual joystick baseline implemented using Unity’s default animation system, where predefined walk and run clips are triggered based on input direction and speed.
The results showed that TouchWalker improved users’ sense of embodiment, enjoyment, and immersion. It also enabled more consistent and accurate control in tasks requiring precise timing and spatial placement, such as stepping on discrete targets. However, it was less effective in fast-paced or spatially constrained scenarios, likely due to the added effort of coordinating finger inputs.
Operating in real time, TouchWalker-MotionNet combined with TouchWalker-UI—a rhythm-based, embodied input interface—supports expressive full-body avatar control and strengthens the user-avatar connection through the immediacy of touch.
}

Our main contributions are as follows:
\begin{itemize}
{\color{rv}
    \item We propose TouchWalker-MotionNet, a real-time neural motion generator that synthesizes full-body avatar locomotion from temporally sparse two-finger input.

    \item We develop TouchWalker-UI, a compact finger-walking interface that enables responsive, per-frame avatar control on touchscreens.

    \item We reconceptualize finger-walking as a continuous and expressive input modality—shifting from symbolic gesture triggering to real-time, embodied avatar control.

    \item We conduct a user study comparing TouchWalker to a virtual joystick baseline with predefined motions, demonstrating improved embodiment, enjoyment, and immersion, and provide design insights through analysis of qualitative feedback and control trade-offs.

}
\end{itemize}

\section{Related Work}

\subsection{Gesture and Touch-Based Locomotion Interfaces}

Various input methods have been proposed to support user locomotion in virtual environments. Among them, walking-based approaches vary widely in how they interpret user input and translate it into movement.
Hung et al.\ introduced camera-based systems such as Puppeteer and FingerPuppet, which use finger gestures to control avatar movement in VR environments \cite{hung_puppeteer_2022, hung_fingerpuppet_2024}. Puppeteer recognizes symbolic gestures—such as walking or jumping—via RGB input and maps them to predefined avatar actions. FingerPuppet extends this concept by retargeting 3D finger trajectories to lower-body motion using handcrafted kinematic rules, and augments them with pre-recorded upper-body clips to produce full-body animation. While both systems visualize avatars, they focus on symbolic gesture interpretation or example-based matching and are not designed for continuous, per-frame motion generation directly from real-time tactile input.
Other studies have explored locomotion control using either finger gestures or real-foot input. FWIP simulates camera translation in VR based on in-place finger gestures performed on a touchscreen \cite{FWIP}. Separately, Lee et al.\ proposed a system that classifies walking gestures from foot pressure using a carpet-type tactile sensor and a vision transformer \cite{FootPad}. These systems focus on controlling user position in virtual space, without generating full-body avatar motion.

Gesture-driven interfaces have also supported trajectory-based locomotion through sketch or mid-air input. Double Doodles allows users to draw motion paths using 6-DOF tracked controllers in immersive VR \cite{DoubleDoodles}, while ARAnimator enables in-situ mobile AR animation through mid-air gestures \cite{ARAnimator}. These systems generate visible avatar motion and emphasize expressive authoring, but are generally not designed for compact or real-time interaction.

\textcolor{rv}{In contrast, TouchWalker uniquely combines TouchWalker-UI, a compact touchscreen-based interface, with TouchWalker-MotionNet, a real-time motion generation model, to synthesize full-body avatar motion from sparse finger contact.} Unlike systems based on symbolic gesture recognition or rule-based kinematic mapping, our approach generates motion frame-by-frame using a learned MoE-GRU network. This enables expressive and responsive avatar control on touchscreen devices.

\subsection{Embodiment and Tactile Feedback in Interaction}

Prior work suggests that body-based and tactile interactions can enhance immersion and bodily engagement in virtual environments.
Dourish describes embodiment as a foundational concept in interaction, where cognition is inseparable from bodily experience \cite{dourish_where_2001}.
Kilteni et al.\ further define virtual embodiment as arising from multisensory integration and sensorimotor contingencies \cite{kilteni_sense_2012}.
Sun et al.\ demonstrate that wearable haptic devices with multimodal tactile feedback—such as vibrotactile and thermal stimuli—can significantly enhance immersive virtual experiences by supporting continuous perception and realistic touch interaction \cite{sun_augmented_2022},
while Li et al.\ show that kinesthetic gloves can significantly strengthen user-avatar association \cite{li_haptics-mediated_2024}.

Though TouchWalker does not employ dedicated haptic hardware, it encourages a tactile sense of engagement by combining physical finger contact on the touchscreen with the alignment of both timing and mapped position between finger taps and avatar foot-ground contact. This coordinated sensorimotor mapping reinforces a grounded sensation and bodily involvement, which was frequently reflected in user feedback during our study. Other systems such as HandAvatar have similarly demonstrated how embodied input metaphors can deepen users’ sense of presence and control \cite{HandAvatar}.

\subsection{Learning-based Motion Generation}

Recent advances in learning-based character animation have produced diverse models for motion generation from minimal or abstract input. Regression models—particularly autoregressive architectures that predict motion frame-by-frame based on prior context—have been widely adopted for interactive applications, with techniques such as phase-functioned networks \cite{holden_phase-functioned_2017} and mixture-of-experts (MoE) models \cite{MANN,starke_neural_2021} improving temporal coherence and reducing motion artifacts. In parallel, generative approaches such as VAEs \cite{ling_character_2020, NeMF, tang_rsmt_2023} and diffusion models \cite{cohan_flexible_2024, shafir2024human, tevet2023human} have enabled more expressive motion synthesis conditioned on a variety of inputs, including text and trajectories.

While recent diffusion models offer high-quality and diverse motion outputs, their reliance on full-sequence generation and relatively higher computational demands pose challenges for real-time use. In contrast, \textcolor{rv}{TouchWalker-MotionNet} adopts an autoregressive architecture designed to support responsive, per-frame motion generation from sparse input. The model uses a lightweight MoE-GRU structure with a foot-alignment loss to ensure contact fidelity. This choice reflects a practical tradeoff—favoring control consistency and real-time interactivity over long-range generative expressiveness.

\subsection{Comparison to Closely Related Work}

Among prior research, the method proposed by Lockwood and Singh \cite{lockwood_finger_2012} shares key conceptual similarities with our approach. Their system interprets finger-walking input on a touch-sensitive tabletop display by capturing sequences of contact points in screen-space and mapping them to predefined full-body walk cycles. While effective for symbolic motion triggering, this approach operates on buffered input and is designed around motion templates, which may reduce flexibility in unstructured scenarios. In addition, character movement remains bounded to the screen-space input region, which narrows the scope of spatial interaction.

\textcolor{rv}{In contrast, TouchWalker-UI interprets finger input on a per-frame basis and passes it to TouchWalker-MotionNet, a learned motion generation model.} Each finger contact is mapped to the avatar’s local coordinate system, allowing the character to move through arbitrarily large environments regardless of the input area. Rather than selecting from pre-authored motion clips, \textcolor{rv}{TouchWalker-MotionNet} generates full-body motion continuously from sparse input, adapting to the user’s rhythm and variation in real time.  
\textcolor{rv}{This continuous generation approach allows for nuanced avatar responses to user input, enabling expressive control beyond symbolic gesture mapping. To evaluate this capability, we designed multi-stage tasks requiring precise coordination and real-time responsiveness, going beyond prior work focused on symbolic gesture mapping.}

\section{\textcolor{rv}{TouchWalker-MotionNet}}

\begin{figure*}
    \centering
    \includegraphics[trim=0 120 0 120, clip, width=.9\linewidth]{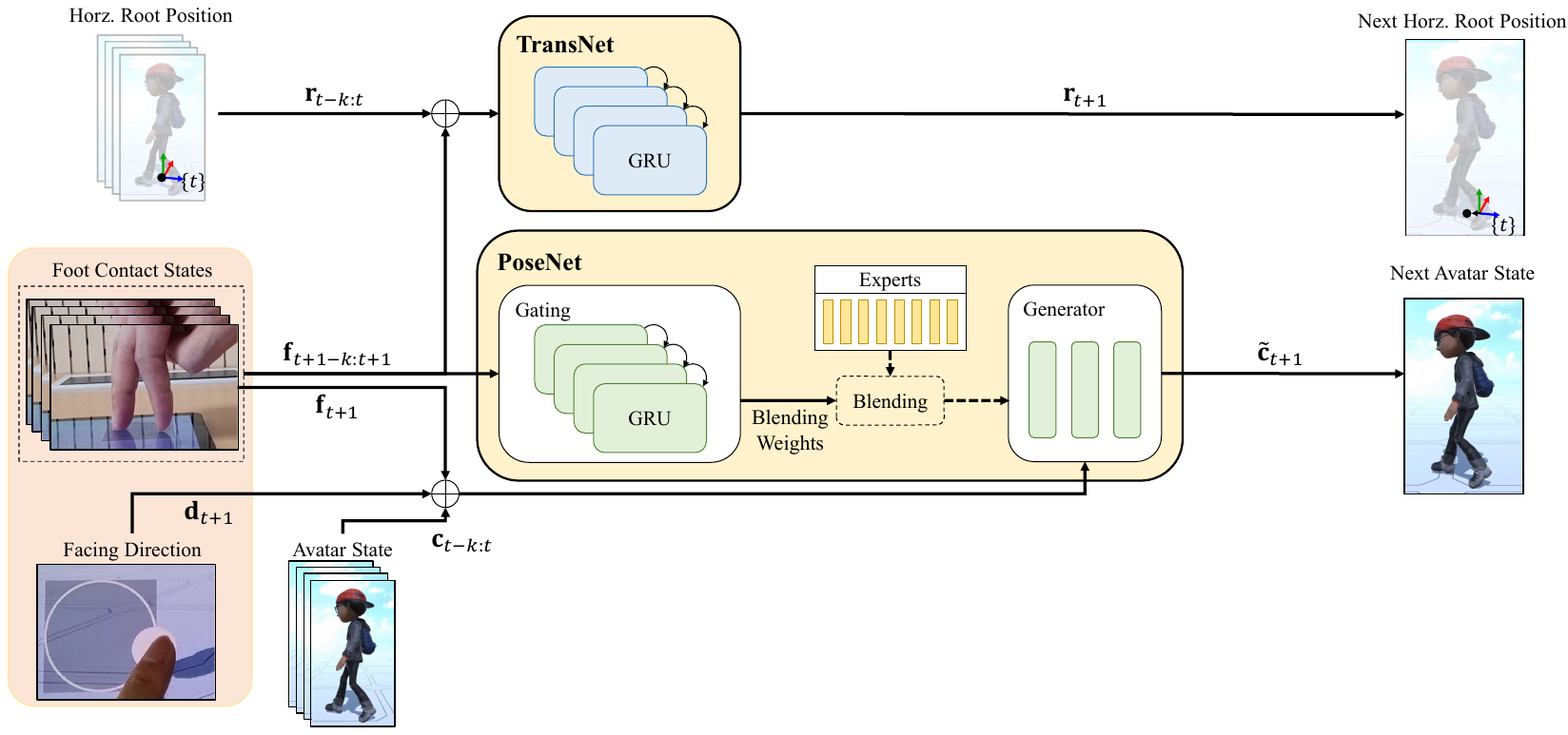}
    \caption{Overview of \textcolor{rv}{TouchWalker-MotionNet}. The model generates full-body avatar motion in real time from touchscreen input, incorporating foot contact positions, facing direction, and state history.}
    \label{fig:overview}
\end{figure*}

As illustrated in Figure~\ref{fig:overview}, \textcolor{rv}{TouchWalker-MotionNet} generates full-body avatar motion in real time from user input on the touchscreen. At each frame, the model receives the user-specified foot contact positions and desired facing direction for the next frame, along with a short history of the avatar's past state, root trajectory, and contact information. These histories are computed from the sequence of previous user inputs and the avatar poses generated so far.

\subsection{Input and Output Representation}

\textbf{Input} $\mathbf x_t$ at current frame $t$ is defined:
\begin{equation} 
    \mathbf x_t = \{ \mathbf f_{t+1-k:t+1}, \mathbf d_{t+1}, \mathbf r_{t-k:t}, \mathbf c_{t-k:t} \}.
    \label{eq:network_input}
\end{equation}

The sequence \(\mathbf{f}_{t+1-k:t+1}\) represents the foot contact states over the most recent \(k+1\) frames, including the current one.
The user's foot contact input at the current frame is used as the contact states for the next frame, \(\mathbf{f}_{t+1}\).
The foot contact states \(\mathbf{f}_{\tau}\) at an arbitrary frame \(\tau\) are defined:
\begin{equation} 
    \mathbf f_{\tau} = \{\mathbf f_{\tau}^p, \mathbf f_{\tau}^v, \mathbf f_{\tau}^c\},
    \label{eq:foot_contact_states}
\end{equation}

where $\mathbf f_{\tau}^p, \mathbf f_{\tau}^v \in \mathbb{R}^4$ denote the horizontal positions and velocities of both feet, expressed in the avatar's local coordinate system at frame ${\tau}$. This frame ${{\tau}}$ is defined with the x-axis from the left hip to the right hip projected onto the ground, the y-axis as the global up direction, and the z-axis as their cross product; the root (pelvis) projection serves as the origin.
$\mathbf f_{\tau}^c \in \mathbb{R}^2$ indicates binary contact labels for the feet. All values in $\mathbf f_{\tau}^p, \mathbf f_{\tau}^v, \mathbf f_{\tau}^c$ are zeroed for the swing foot.

The vector $\mathbf{d}_{t+1} \in \mathbb{R}^2$ denotes the user-specified facing direction for the next frame, expressed relative to the avatar's current facing at frame $t$. The angular difference is encoded as a 2D vector using sine and cosine.

The sequence \(\mathbf{r}_{t-k:t}\) represents the 2D root positions over the most recent \(k+1\) frames, all expressed in the avatar's local coordinate system at current frame $t$, \(\{t\}\).

The sequence $\mathbf c_{t-k:t}$ represents the avatar state over the most recent $k+1$ frames,
which is defined as:
\begin{equation} 
    \mathbf c_{t-k:t} = \{\mathbf o_{t-k:t}, \mathbf p_{t-k:t}, \mathbf v_{t-k:t} \},
    \label{eq:character_state}
\end{equation}
where $\mathbf o_{t-k:t} \in \mathbb{R}^{J \times 6 \times k}, \mathbf p_{t-k:t}, \mathbf v_{t-k:t} \in \mathbb{R}^{J \times 3 \times k}$ represent 6D orientations, positions, velocities of all joints over $k+1$ frames where $J$ is the number of joints,
all expressed in \(\{t\}\).

\textbf{Output} $\mathbf{y_t}$ is defined as:
\begin{equation} 
    \mathbf y_t = \{ \mathbf r_{t+1}, \tilde{\mathbf c}_{t+1} \},
    \label{eq:network_output}
\end{equation}
where $\mathbf r_{t+1} \in \mathbb{R}^{2}$ represents the 2D root position at the next frame expressed in $\{t\}$,
and \(\tilde{\mathbf{c}}_{t+1}\) is defined as the reduced avatar state for the next frame, as follows:
\begin{equation} 
    \tilde{\mathbf c}_{t+1} = \{ \mathbf o_{t+1}, h_{t+1}, \mathbf v_{t+1} \},
    \label{eq:reduced_character_state}
\end{equation}
where $\mathbf o_{t+1}$ and $\mathbf v_{t+1}$ represent 6D orientations and velocities of all joints at the next frame expressed in $\{t\}$,
and $h_{t+1}$ represents the global root height.

\subsection{Network Architecture}

\textcolor{rv}{TouchWalker-MotionNet} comprises two main components: TransNet and PoseNet.

TransNet takes as input the horizontal root positions $\mathbf r_{t-k:t}$ over the past $k{+}1$ frames and the foot contact states $\mathbf f_{t+1-k:t+1}$ spanning the past $k$ frames and the next frame, and outputs the root position $\mathbf{r}_{t+1}$ for the next frame. This design reflects the observation that root translation is more directly influenced by the recent motion of the root and stance foot than by pose or facing direction. All root positions are expressed relative to the current avatar coordinate system $\{t\}$. To model the temporal structure, we use a GRU-based network with two layers (hidden size 32), processing $k{+}1$ time steps. We set $k{=}5$ in all experiments.

PoseNet receives $\mathbf f_{t+1-k:t+1}$, $\mathbf c_{t-k:t}$, and $\mathbf d_{t+1}$ as input, and predicts the next-frame avatar state $\tilde{\mathbf c}_{t+1}$. To address the mean pose issue, PoseNet adopts a Mixture of Experts (MoE) architecture inspired by \cite{MANN}, but it incorporates GRU layers in the gating network to leverage temporal context even without continuous input. The gating network takes $\mathbf f_{t+1-k:t+1}$ and outputs blending weights for $K$ expert networks. These weights are used to blend expert parameters, forming a generator network that produces $\tilde{\mathbf c}_{t+1}$ based on $\mathbf f_{t+1}$, $\mathbf d_{t+1}$, and $\mathbf c_{t-k:t}$. The gating network consists of one GRU layer (hidden size 32), and each expert network has two fully connected layers of size 128 with ELU activations. We set $K{=}8$ in all experiments.

\subsection{Loss Function}

The loss function for training is defined as:
\begin{equation} 
    \label{total_loss}
    \begin{split}
        L = & w_1 \cdot L_\mathrm{rec} + w_2 \cdot {L}_\mathrm{FK} + w_3 \cdot {L}_\mathrm{dir} +\\
            & w_4 \cdot {L}_\mathrm{ct} + w_5 \cdot {L}_\mathrm{ct\_trans},
    \end{split}
\end{equation}
where $w_\text{1-5}$ are the weights for the loss terms, and the reconstruction loss $L_\mathrm{rec}$ is defined as:
\begin{equation} 
    \label{eq:L_rec}
    L_\textrm{rec} = \textrm{MSE}(\tilde{\mathbf c}_{t+1}, \hat{\tilde{\mathbf c}}_{t+1}),
\end{equation}
where $\tilde{\mathbf c}_{t+1}$ and $\hat{\tilde{\mathbf c}}_{t+1}$ represent the ground truth avatar state and its predicted value, respectively.

The FK loss $L_\mathrm{FK}$ is defined as follows:
\begin{equation} 
    \label{eq:L_FK}
    L_\textrm{FK} = \textrm{MSE}(\mathbf p_{t+1}, \textrm{FK}(\hat{\mathbf r}_{t+1}, \hat{h}_{t+1}, \hat{\mathbf o}_{t+1})),
\end{equation}
where $\mathbf p_{t+1}$ refers to the ground truth joint positions and $\hat{\mathbf r}_{t+1}$, $\hat{h}_{t+1}$, and $\hat{\mathbf o}_{t+1}$ refer to the predicted 2D root position, root height, and joint orientations, respectively.
$\textrm{FK}(\cdot)$ represents a FK layer~\cite{FKLayer} which converts the predicted values into joint positions.

The direction loss $L_\textrm{dir}$ is defined as:
\begin{equation} 
    \label{eq:L_dir}
    L_\textrm{dir} = \textrm{MSE}(\textrm{log}(\mathbf T_{t}^{-1}\mathbf T_{t+1}), \textrm{log}(\mathbf D_{t+1})),
\end{equation}
where $\mathbf T_{t}, \mathbf T_{t+1} \in \textrm{SE(3)}$ represent the avatar coordinates at frame $t$ and $t+1$, 
and $\mathbf D_{t+1} \in \textrm{SO(3)}$ is the expansion of $\mathbf d_{t+1}$ into a 3D rotation matrix that signifies rotation around the global vertical axis.
$\textrm{log}(\cdot)$ converts a rotation matrix into a rotation vector.

Inspired by \cite{FootContactLoss}, the contact loss $L_\textrm{ct}$ is calculated as:
\begin{equation} 
    \label{eq:L_contact}
    L_\textrm{ct} = \| (\mathbf f_t^c \odot \mathbf f_{t+1}^c) \odot \textrm{MSE}_{\textrm{feet}}(\mathbf f^\textrm{FK}_{t+1}, \mathbf f_t^p) \|_1,
\end{equation}
where $\odot$ is a element-wise multiplication operator and $\textrm{MSE}_{\textrm{feet}}$ refers to a function that takes two 4D vectors, each representing the horizontal positions of two feet, and returns a 2D vector containing the MSE for each foot.
$\mathbf f^\textrm{FK}_{t+1}$ represents the predicted horizontal positions of two feet at $t+1$ with respect to $\{t\}$, obtained via the FK layer.
$\mathbf f^c_t$ and $\mathbf f^p_t$ represent the binary contact labels and horizontal positions of two feet at $t$.
This loss term encourages PoseNet and TransNet to produce outputs $\hat{\mathbf r}_{t+1}$, $\hat{h}_{t+1}$, and $\hat{\mathbf o}_{t+1}$ such that the foot positions at $t+1$, calculated from $\hat{\mathbf r}_{t+1}$, $\hat{h}_{t+1}$, and $\hat{\mathbf o}_{t+1}$, are the same as the foot positions at $t$, if the foot maintains contact across two adjacent frames.
However, in practice, even if the predicted next root position is inaccurate, as long as the positions of the feet match through unexpected rotations of other joints, $L_\textrm{ct}$ is satisfied, leading to incorrect predictions of $\hat{\mathbf r}_{t+1}$.

Hence, we employ an additional contact transform loss, $L_\textrm{ct\_trans}$, focused on accurately predicting $\hat{\mathbf r}_{t+1}$ and the root orientation in $\hat{\mathbf o}_{t+1}$ while contact is maintained.
This loss is calculated as:
\begin{equation} 
    \label{eq:L_contact_transform}
    L_\textrm{ct\_trans} = \| (\mathbf f_t^c \odot \mathbf f_{t+1}^c) \odot \textrm{MSE}_{\textrm{feet}}(\mathbf f_{t+1}^{p\{t\}}, \mathbf f_t^p) \|_1,
\end{equation}
where $\mathbf f_{t+1}^{p\{t\}}$ represents the feet positions at $t+1$ with respect to $\{t\}$, not calculated from the network output, but transformed from the network input $\mathbf f_{t+1}^p$ which is originally expressed in $\{t+1\}$.
This transformation is derived from the hip joint positions transformed by $\hat{\mathbf r}_{t+1}$ and the root orientation in $\hat{\mathbf o}_{t+1}$.
This loss term encourages TransNet and PoseNet to output them such that the foot positions $\mathbf f_{t+1}^{p\{t\}}$ at $t+1$, transformed from $\mathbf f_{t+1}^p$, are the same as the foot positions $\mathbf f_t^p$ at $t$, if the foot maintains contact.
This loss has a similar meaning to how $L_\textrm{FK}$ and $L_\textrm{rec}$ ensure that the root's 2D position and orientation match the ground truth.
However, this loss differs in that it only applies based on the foot's position on the horizontal plane when the foot's contact is maintained.

\subsection{Training}

Training is conducted end-to-end, and the gating network and the $K$ expert networks of PoseNet and TransNet are trained simultaneously.
We train and evaluate the model on approximately 100 minutes of motion data from the LaFAN1 dataset~\cite{LaFAN1}.
Foot contact states and facing directions are extracted from the motion data. A foot is labeled as in contact if its height and speed fall below \SI{0.1}{m} and \SI{0.24}{m/s}, respectively. These labels are also used to construct $\mathbf f_{\tau}^c$.
Training details are provided in Section~1 of the supplementary document.

\subsection{Runtime}

At each frame, the system receives user input in the form of finger contact and facing direction. The finger tap is interpreted as a stance foot position, from which we construct the foot contact state vector $\mathbf{f}_{t+1}$.
For the touched (stance) foot, the position $\mathbf f^p_{t+1}$ is set to the touch position, velocity $\mathbf f^v_{t+1}$ is computed by finite difference from the previous frame, and contact flag $\mathbf f^c_{t+1}$ is set to 1. For the swing foot, all values are set to zero.
The facing direction $\mathbf{d}_{t+1}$ is provided via a virtual joystick and encoded as a 2D vector relative to the current heading.

The model generates the next-frame avatar pose based on the past $k+1$ frames of avatar state, root trajectory, and foot contact. 
A smoothing filter is applied to reduce temporal jitter.
To handle uneven terrain, the model’s predicted foot and root heights are added to the sampled terrain height at their respective horizontal positions. The avatar's lower-body pose is then refined via inverse kinematics to match the adjusted positions.
The runtime system runs on-device at nearly 30 FPS (Galaxy Tab S7+); see Section~2 in the supplementary document for runtime implementation and performance details.

\section{\textcolor{rv}{TouchWalker-UI}}

\begin{figure}
  \centering
  \subfloat[]{
      \includegraphics[trim=200 45 350 160, clip, width=.49\linewidth]{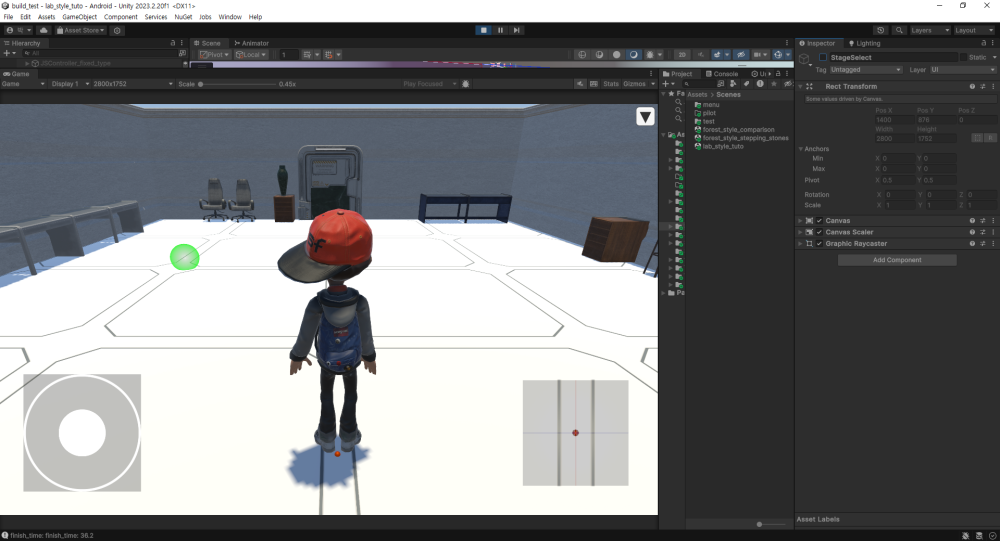}
      \label{fig:touchwalker-interface}}%
  \subfloat[]{
      \includegraphics[trim=170 20 120 20, clip, width=.49\linewidth]{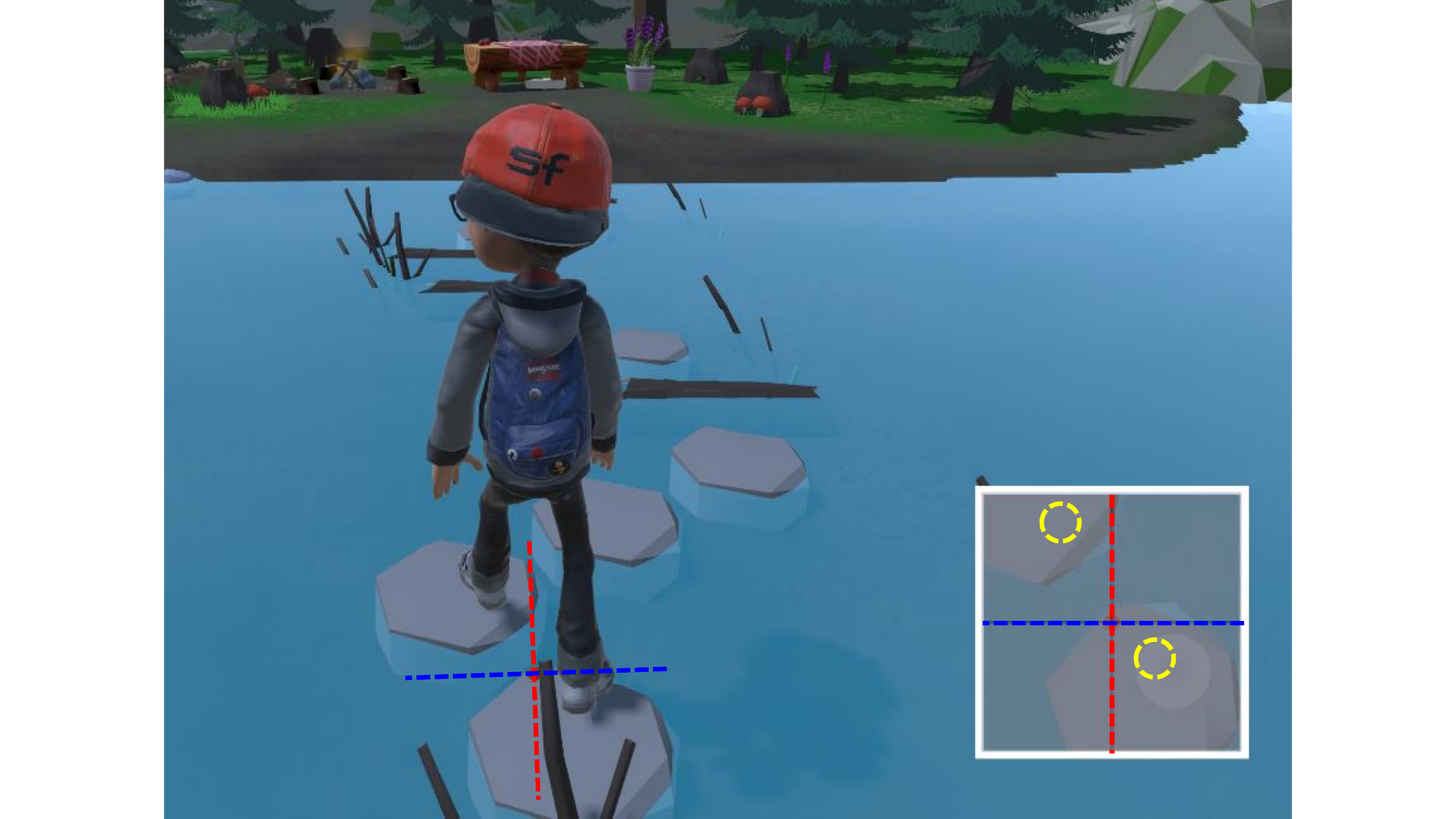}
      \label{fig:touchwalker-example}}%
  \caption{
  (a) \textcolor{rv}{TouchWalker-UI}. The touch region (bottom-right square) displays a top-down rendering of the ground beneath the avatar. (b) The touched position (yellow circle) is interpreted relative to the avatar’s horizontal position and orientation on the ground. 
  The red and blue dotted lines indicate the avatar’s forward and lateral directions in both the 3D virtual space and the touch region. All visual 
  markers are added to the figure for illustration only.
  }
\label{fig:user-study-methods}
\end{figure}

\textcolor{rv}{TouchWalker-UI} provides a real-time locomotion interface that enables users to control a full-body avatar using two-finger gestures on a touchscreen. \textcolor{rv}{It implements} a simple and intuitive multitouch input paradigm that mimics the act of walking, while allowing users to guide avatar motion continuously and responsively \textcolor{rv}{on top of the TouchWalker-MotionNet}.

Users \textcolor{rv}{interact with TouchWalker-UI} by rhythmically alternating two fingers within a designated touch region displayed on the screen.
This region displays a top-down rendering of the ground beneath the avatar, allowing users to visually infer where to place each foot relative to the avatar's position and orientation (Figure~\ref{fig:touchwalker-interface}).

{\color{rv}
Each touch is interpreted as the contact point of a foot. The first touch on the left or right side of the region is mapped to the corresponding foot. If a second touch appears while one foot is already assigned, it is mapped to the other foot. These assignments persist until fingers are lifted. When only one finger remains, the unmapped foot is considered to be in swing phase.

The stance foot position—interpreted relative to the avatar’s horizontal position and orientation—is passed to \textcolor{rv}{TouchWalker-MotionNet} as input at every frame. To support intuitive alignment, the touch region includes visual indicators: a vertical line showing the avatar’s forward direction, a horizontal line for lateral alignment, and a red dot marking the avatar’s center. These cues help users align their touches to the avatar’s frame in the 3D space (Figure~\ref{fig:touchwalker-example}).

While \textcolor{rv}{TouchWalker-UI} enables avatar-relative movement control, it does not directly allow users to rotate the avatar. To address this, we include a separate virtual joystick for facing direction. Rather than rotating the generated motion post hoc—which risks artifacts like foot sliding—we pass the desired facing direction as input to \textcolor{rv}{TouchWalker-MotionNet}, enabling it to produce full-body poses that are coherent with both foot placement and intended facing.}

\section{User Study Design}

\textcolor{rv}{To evaluate the effectiveness of our proposed TouchWalker system, we conducted a user study comparing two control methods for avatar control: (i) TouchWalker, which uses two-finger touch input to generate full-body motion in real time via a neural network, and (ii) a virtual joystick, which controls avatar movement by replaying predefined animation clips.}
Participants completed two tasks requiring an avatar to navigate through various environmental obstacles to reach a designated goal. Task performance was recorded during and after the tasks, and participants completed a questionnaire assessing their sense of 
embodiment, enjoyment, and immersion.
Additionally, qualitative feedback was collected through post-task interviews.

\subsection{Hypotheses}

\textcolor{rv}{Based on prior findings that body-based input and tactile cues can enhance user engagement and immersion \cite{kilteni_sense_2012, dourish_where_2001,sun_augmented_2022,li_haptics-mediated_2024}, and the design of TouchWalker—which combines TouchWalker-UI for tactile foot-ground interaction and TouchWalker-MotionNet for responsive full-body motion—we hypothesized that the system would lead to higher ratings of embodiment, enjoyment, and immersion compared to virtual joysticks.}
Furthermore, we explored whether prior experience with joystick controls would influence users’ subjective experience.
This led us to formulate the following hypotheses:
 
\begin{description}
\item[H1.] TouchWalker will lead to higher ratings of embodiment, enjoyment, and immersion compared to the virtual joystick.

\item[H2.] The difference in subjective ratings between TouchWalker and the virtual joystick will be influenced by  participants’ prior experience with virtual joystick controls.
\end{description}

\subsection{Participants}

A total of 14 participants (9 males, 5 females), aged between 18 and 39 were recruited from the university community following approval from the university’s IRB.
According to a preliminary questionnaire, participants reported varying degrees of familiarity with virtual joystick controls on a 5-point scale. Responses ranged from 1 ("Not at all familiar") to 5 ("Very familiar"), with 2 participants rating themselves as 1, 2 as 2, 5 as 3, 3 as 4, and 2 as 5.
One participant was left-handed, while the remaining thirteen were right-handed.

\subsection{Materials}

\begin{figure}
  \centering
  \subfloat[T1-S1: Hole Avoidance]{
      \includegraphics[trim=20 30 225 190, clip, width=.32\linewidth]{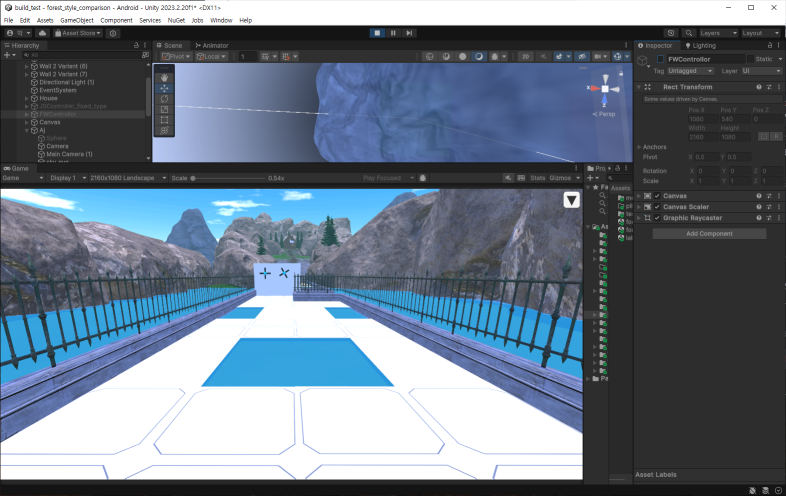}
      \label{fig:t1-s1}}%
  \subfloat[T1-S2: Speed Control]{
      \includegraphics[trim=20 30 225 190, clip, width=.32\linewidth]{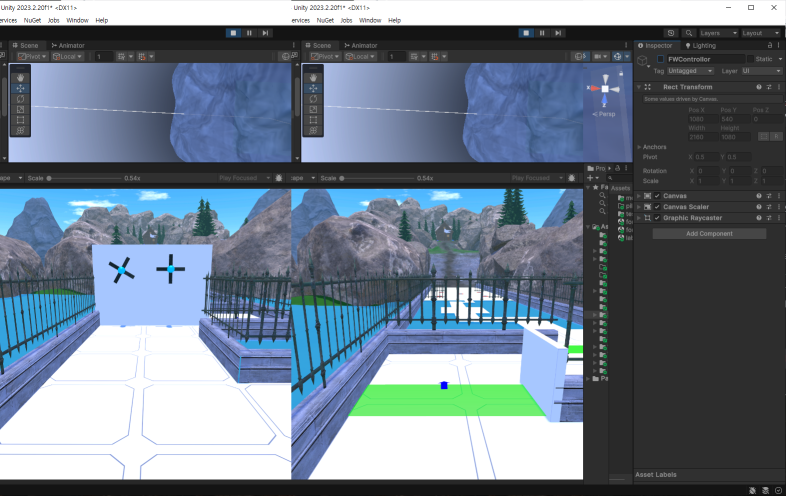}
      \label{fig:t1-s2}}%
  \subfloat[T1-S3: Narrow Path]{
      \includegraphics[trim=20 30 225 190, clip, width=.32\linewidth]{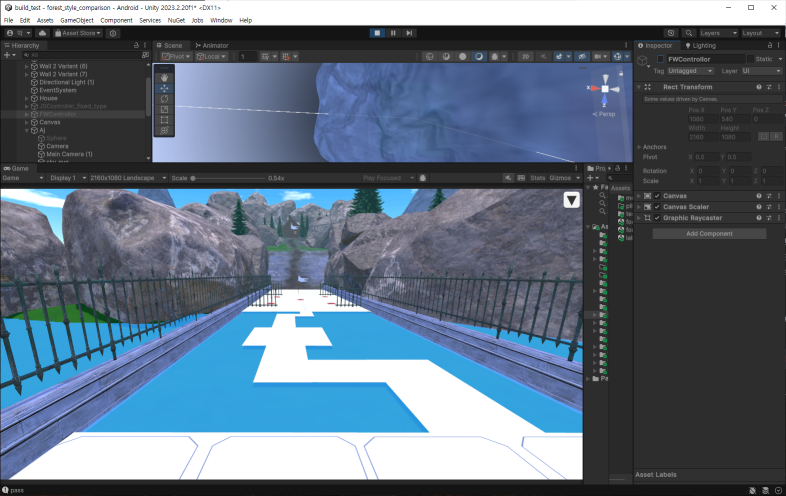}
      \label{fig:t1-s3}}%
  \\
  \subfloat[T1-S4: Foot Buttons]{
      \includegraphics[trim=20 30 225 190, clip, width=.32\linewidth]{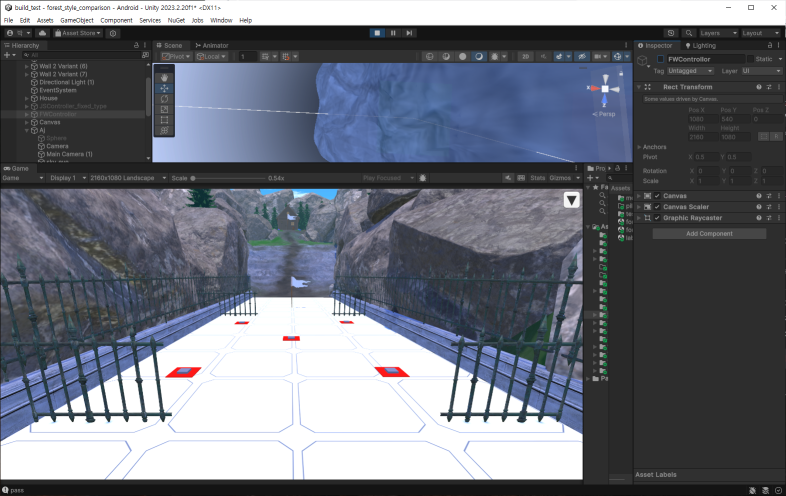}
      \label{fig:t1-s4}}%
  \subfloat[T1-S5: Rolling Obstacles]{
      \includegraphics[trim=20 30 225 190, clip, width=.32\linewidth]{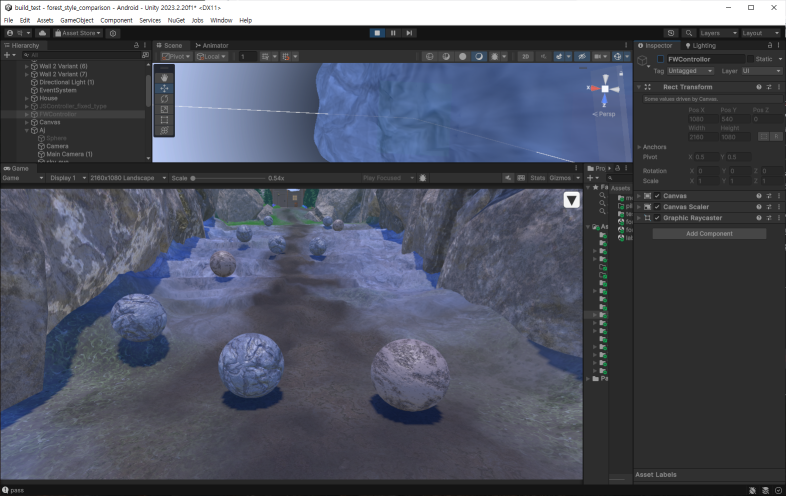}
      \label{fig:t1-s5}}%
  \subfloat[T2: Stepping Stones]{
      \includegraphics[trim=20 30 225 190, clip, width=.32\linewidth]{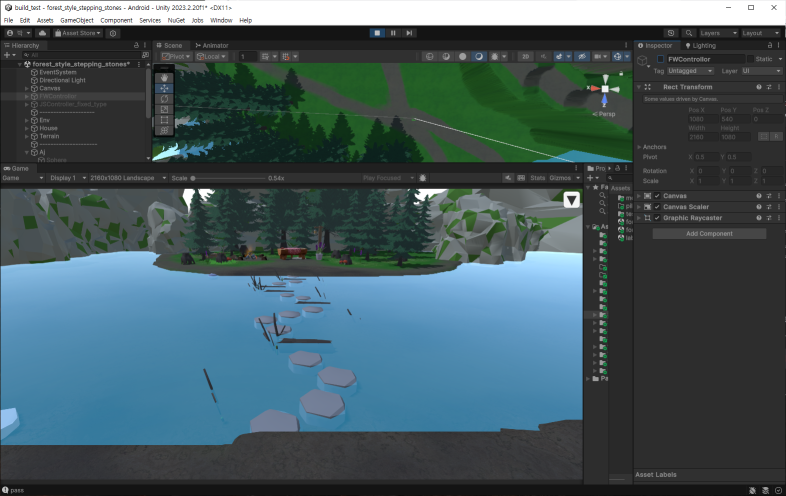}
      \label{fig:t2}}%
  \caption{
    Scenes from the user study tasks. T1–S1 to T1–S5 denote the five stages of Task 1 (Multi-Stage Navigation); T2 denotes Task 2 (Stepping Stones).
  }
\label{fig:user-study-scenes}
\end{figure}

Participants completed two tasks on an Android tablet (Galaxy Tab S7+), each involving a test environment developed with Unity.
Each task was designed to assess various aspects of movement control, such as speed modulation, precision, quick adjustments, and the control of foot placement and contact duration (Figure~\ref{fig:user-study-scenes}).

    \noindent\textbf{Task 1 (Multi-stage Navigation)}:
    The experimental environment for this task was designed as a game-like setting, where participants controlled an avatar to navigate through multiple stages, following a style similar to that proposed in \cite{lee_novel_2024}. 
    The environment consisted of the following five stages:
        
        \noindent\textbf{Stage 1 (Hole Avoidance)}
        with holes scattered along the path, where falling results in returning to the start, requiring basic movement control.
        
        \noindent\textbf{Stage 2 (Speed Control)}
        composed of a strong wind part and a speed-limit part where participants must move above a certain speed to overcome the wind, and below a certain speed to pass through speed-limit zones. This stage requires control over varying movement speeds.
        
        \noindent\textbf{Stage 3 (Narrow Path)}
        with a narrow and winding path, surrounded by holes that return the avatar to the start if fallen into, requiring precise movement control.
        
        \noindent\textbf{Stage 4 (Foot Buttons)}
        with five foot buttons that must be activated one by one by stepping on each for one second. To proceed, all five buttons must be activated, requiring control over avatar and foot placement.
        Both control methods can perform this task, but virtual joystick users indirectly press the buttons by positioning the avatar over them, while the TouchWalker allows direct control of foot contact through touch input.
        
        \noindent\textbf{Stage 5 (Rolling Obstacles)}
        with rolling obstacles coming down a slope that must be avoided while reaching the goal. Getting hit bounces the avatar backward, so participants must avoid collisions and reach the goal, requiring quick and precise control.
    
    \noindent\textbf{Task 2 (Stepping Stones)}:
    with scattered stones and wooden planks as the only traversable surfaces, requiring participants to cross the river without touching the water.
    Wooden planks gradually sink upon contact, encouraging quick transitions to the next foothold.
    This task was specifically designed for the TouchWalker, which allows control over both foot placement and contact timing.
    Although virtual joystick control is not well-suited for this type of interaction, we included it to evaluate how effectively finger walking can prevent water contact by directly controlling foot contact position and duration, in contrast to joystick-based control which only modulates movement velocity.

\begin{figure}
  \centering
  \subfloat[TouchWalker (TW)]{
      \includegraphics[trim=0 30 344 130, clip, width=.49\linewidth]{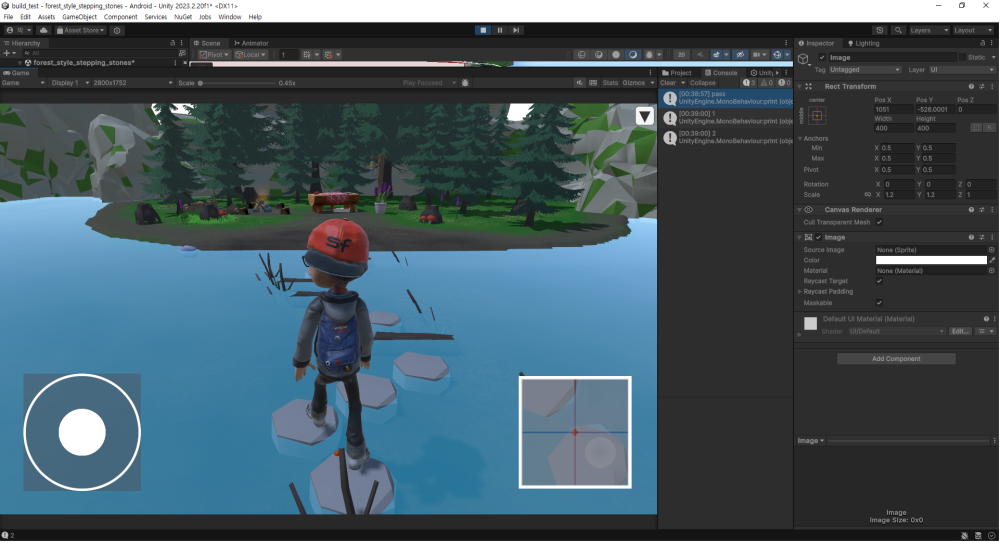}
      \label{fig:control-tw}}%
  \subfloat[Virtual Joystick (VJ)]{
      \includegraphics[trim=0 30 344 130, clip, width=.49\linewidth]{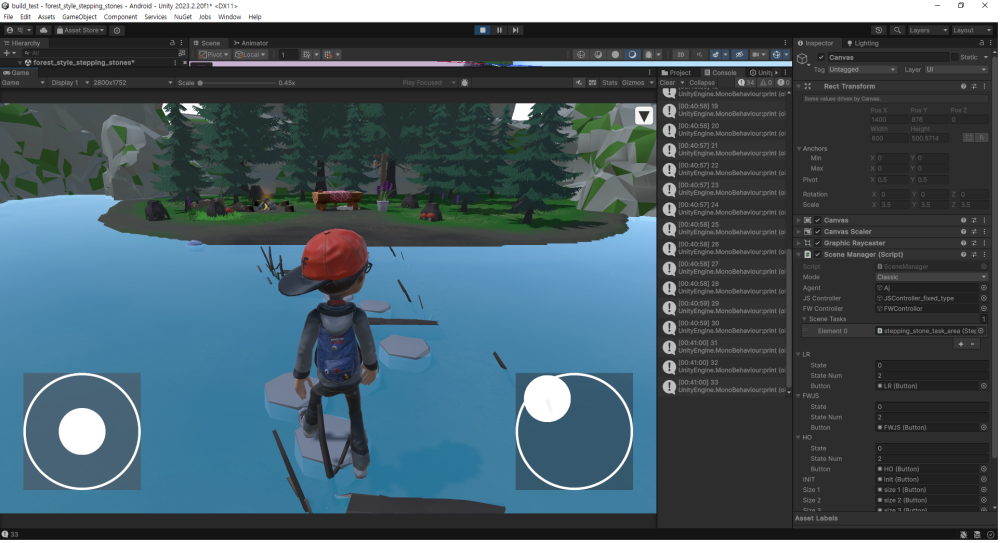}
      \label{fig:control-vj}}%
  \caption{
  Two control methods compared in the user study.
  }
\label{fig:user-study-methods}
\end{figure}

Participants performed each task using the two control methods described below. In both methods, a virtual joystick positioned on the non-dominant-hand side was used to control the avatar’s facing direction. The key difference lies in the design of the movement input, which was placed on the dominant-hand side.

\noindent\textbf{TouchWalker (TW)}:
In our proposed method, the movement of the avatar is controlled via a finger-walking input area located on the dominant-hand side (Figure~\ref{fig:control-tw}).
This contrasts with the standard dual-analog control scheme, where the left thumb (non-dominant hand for right-handed players) controls character movement with relatively low precision, while the right thumb (dominant hand) handles high-precision actions such as orientation control typically associated with weapon aiming \cite{teather_tilt-touch_2017}.  In our interface, however, precise control over locomotion is more critical than facing direction. Therefore, assigning the finger-walking input to the dominant hand is consistent with established principles of manual asymmetry, which suggest allocating fine motor tasks to the dominant hand \cite{guiard_asymmetric_1987}.
\textcolor{rv}{In this method, each finger tap is interpreted in real time by a neural motion generation model that synthesizes full-body avatar motion on a per-frame basis.
This setup enables frame-by-frame motion generation that reflects finger input in both timing and contact location.}

\noindent\textbf{Virtual Joystick (VJ)}:
As a comparison baseline, we placed a virtual joystick for movement control on the dominant-hand side (Figure~\ref{fig:control-vj}). In mobile games, touch interfaces are commonly designed to simulate physical gamepad inputs using on-screen joysticks and buttons \cite{teather_tilt-touch_2017, baxter_virtual_2023}. Among these, our configuration follows the dual-analog virtual joystick style, where one joystick controls movement and the other controls avatar's facing direction. To ensure a fair comparison with the TouchWalker, we mirrored the control layout by assigning movement to the dominant hand and facing direction to the non-dominant hand.
\textcolor{rv}{In this method, avatar motion is driven by Unity’s default animation system, which plays predefined walk or run clips based on joystick direction and speed, without adapting to input rhythm or foot placement.}

\subsection{Procedure}

\begin{figure}
  \centering
  \subfloat[]{
      \includegraphics[trim=0 30 344 130, clip, width=0.32\linewidth]{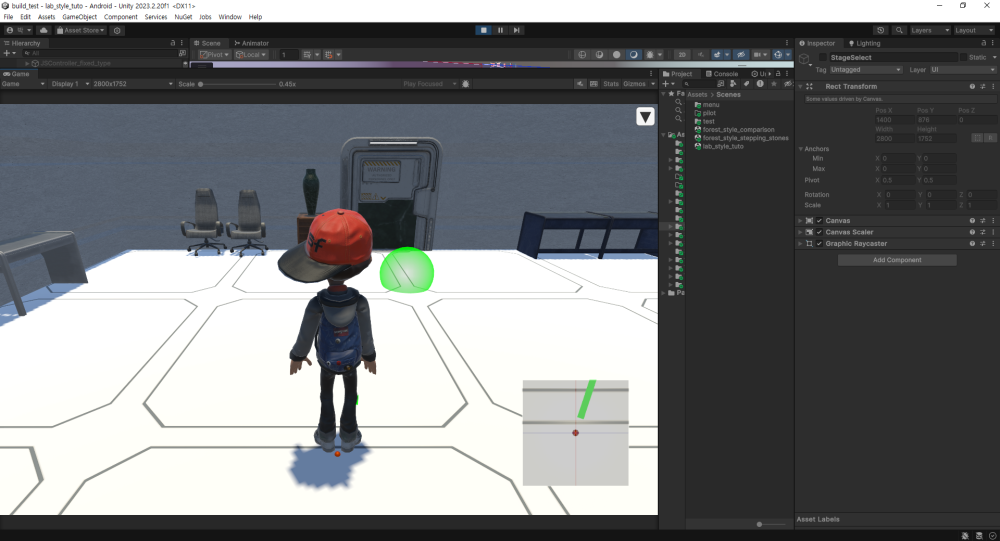}
      \label{fig:tutorial1}}%
  \subfloat[]{
      \includegraphics[trim=0 30 344 130, clip, width=.32\linewidth]{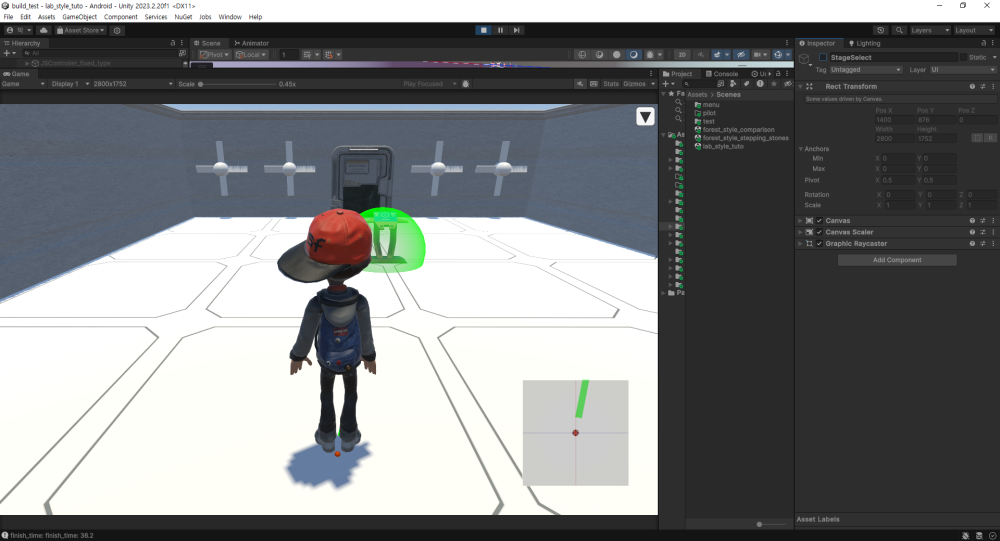}
      \label{fig:tutorial2}}%
  \subfloat[]{
      \includegraphics[trim=0 30 344 130, clip, width=.32\linewidth]{figs/tuto_stage3.png}
      \label{fig:tutorial3}}%
  \caption{
Tutorial stages used in the user study.  
(a) Movement Direction Control: Participants navigated toward sequential green goal spheres using only movement input, with facing direction fixed. An arrow below the avatar indicated the off-screen goal direction.  
(b) Movement Speed Control: Participants adjusted movement speed to reach a forward goal while being pushed back, learning to modulate speed without facing control.  
(c) Facing Direction Control: Participants reached multiple goals without on-screen arrows, practicing both facing and movement control to locate and approach targets.
  }
\label{fig:user-study-tutorial}
\end{figure}

After a verbal overview of the study, participants completed a consent form and a demographic questionnaire, which collected age, gender, dominant hand, and self-reported virtual joystick familiarity. The experimenter also measured the length of the participant’s dominant hand (wrist to middle fingertip). Participants were randomly assigned to either a TW-first or VJ-first task order group with balanced distribution.
For each control method, participants completed a tutorial, performed two main tasks, and answered a post-task questionnaire before moving to the next method.

Tutorials lasted up to 10 minutes and consisted of three stages: (i) direction control, (ii) speed control, and (iii) facing control (Figure~\ref{fig:user-study-tutorial}). Each stage included verbal and on-screen instructions, and participants practiced moving toward goal markers while navigating environmental conditions such as wind or missing visual cues.

In the TW condition, participants additionally selected a preferred square touch area size from five candidates (x0.5–x1.5 relative to a \SI{6}{cm} baseline). The baseline—\SI{4.25}{cm} per side—was empirically determined for comfortable finger spreading. Size selection was done within the tutorial environment and completed within 5 minutes. The selected size was used for all TW tasks.

In Task 1, participants navigated a five-stage obstacle course under time pressure.
Completion time for each stage was recorded, along with stage-specific performance metrics:
number of falls (Stage 1 and 3), duration of speed violations (Stage 2), and number of collisions with obstacles (Stage 5).

Task 2 involved precise stepping: participants crossed a river using only stones and planks, minimizing foot contacts with water. Because this task emphasized deliberate placement over speed, we used only water contacts as the performance metric.

The post-task questionnaire consisted of 11 items selected from established instruments--the Embodiment Questionnaire (EQ) \cite{peck_avatar_2021}, Intrinsic Motivation Inventory (IMI) \cite{ryan_self-determination_2000}, and Immersive Experience Questionnaire (IEQ) \cite{jennett_measuring_2008}--to evaluate users’ sense of embodiment, enjoyment, and immersion.
The full list of items are provided in Section~3 of the supplementary document.
To ensure consistency across all questionnaire items, all items were rated on a unified 7-point Likert scale.
As only selected items were used from each instrument, the results are not intended for comparison with prior studies using the full validated scales,
but rather for within-study comparisons between TW and VJ.
Once both methods were tested, a short interview was conducted to collect qualitative feedback.

\section{Results}

{\color{rv}
All subjective measures were analyzed using non-parametric tests due to the ordinal nature of Likert-scale data. We used Wilcoxon tests to compare paired scores between TW and VJ conditions. To evaluate the influence of prior joystick experience, we performed Mann–Whitney U tests for between-group comparisons, and Spearman's rank correlation to assess associations between joystick familiarity and subjective ratings.

For objective measures (e.g., completion time, number of collisions), we reported descriptive statistics including the median and interquartile range (IQR). Paired t-tests were used for measures that met the normality assumption, while Wilcoxon signed-rank tests were applied otherwise. Outlier removal was not performed, as all recorded values were treated as valid observations reflecting the natural variability in user performance.
Normalization was also not applied, as all comparisons were made within subjects, and the relative differences were considered meaningful without rescaling. No correction for multiple comparisons was applied, following common practice in exploratory user studies of this scale.
}

\subsection{Subjective Evaluation}

\subsubsection{User Ratings}

\paragraph{Section-Level Comparisons}

\begin{figure}
  \centering
  \includegraphics[trim=65 30 65 30, clip,width=\linewidth]{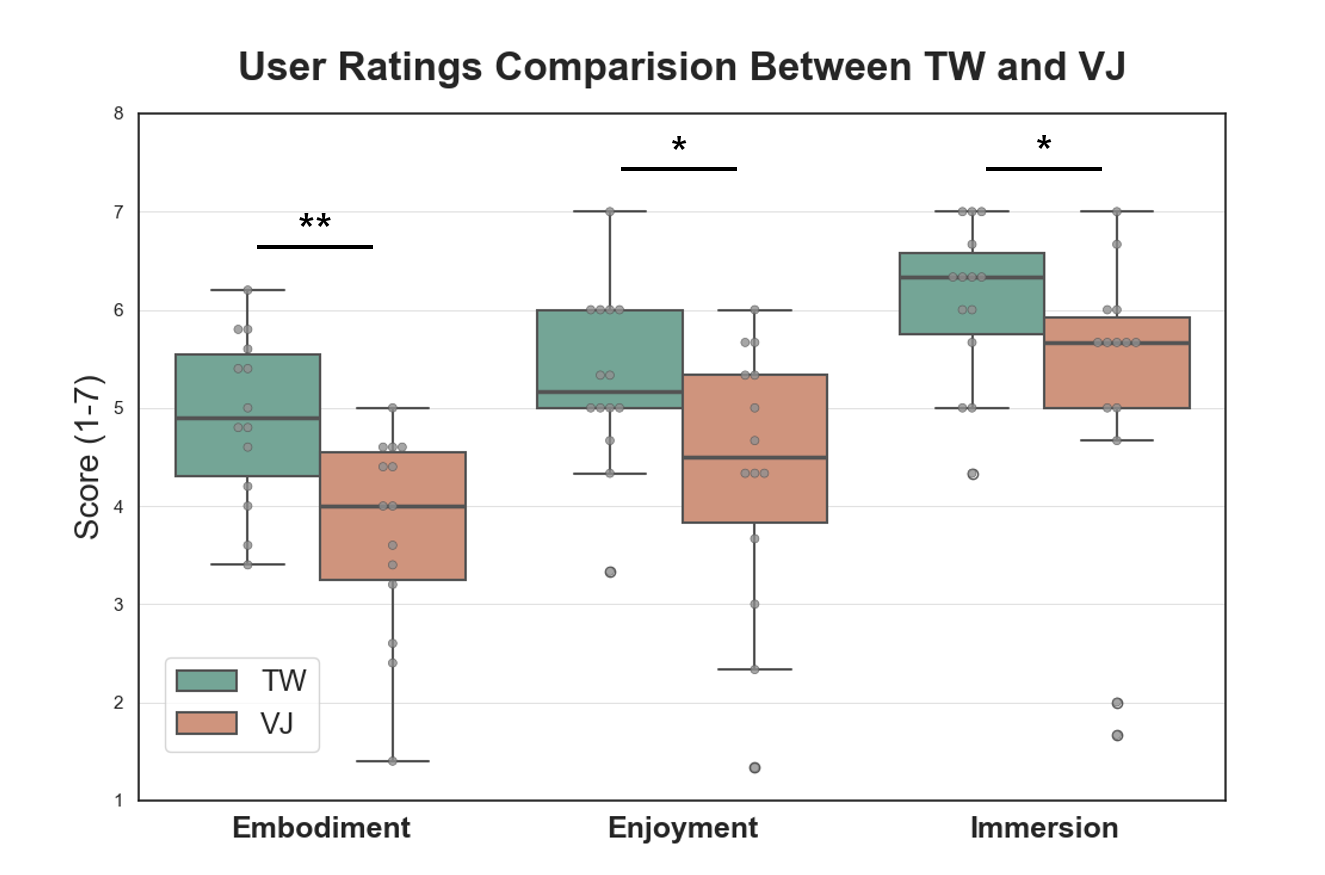}
  \caption{Comparison of TW and VJ on three questionnaire sections. Individual responses are shown as translucent dots. Significant differences were found in all sections (* $p<.05$, ** $p<.01$).}
  \label{fig:section-comparison}
\end{figure}

Wilcoxon signed-rank tests on the section-level averages revealed that TW was rated significantly higher than VJ in both Embodiment ($W = 10.0$, $p = 0.0052$, $r = 0.713$), Enjoyment ($W = 11.5$, $p = 0.0128$, $r = 0.688$), and Immersion ($W = 12.0$, $p = 0.0175$, $r = 0.679$), all indicating large effect sizes (see Figure~\ref{fig:section-comparison}).

In terms of central tendency and variability, TW achieved consistently higher medians and comparable or narrower interquartile ranges across all three dimensions:
Embodiment TW (median = 4.90, IQR = 1.25) vs. VJ (median = 4.00, IQR = 1.30);
Enjoyment TW (5.17, 1.00) vs. VJ (4.50, 1.50);
Immersion TW (6.33, 0.83) vs. VJ (5.67, 0.92).
These results reinforce the consistent and practically meaningful preference for TW between participants.

\paragraph{Influence of Virtual Joystick Experience}

\begin{figure}
  \centering
  \includegraphics[width=1.\linewidth]{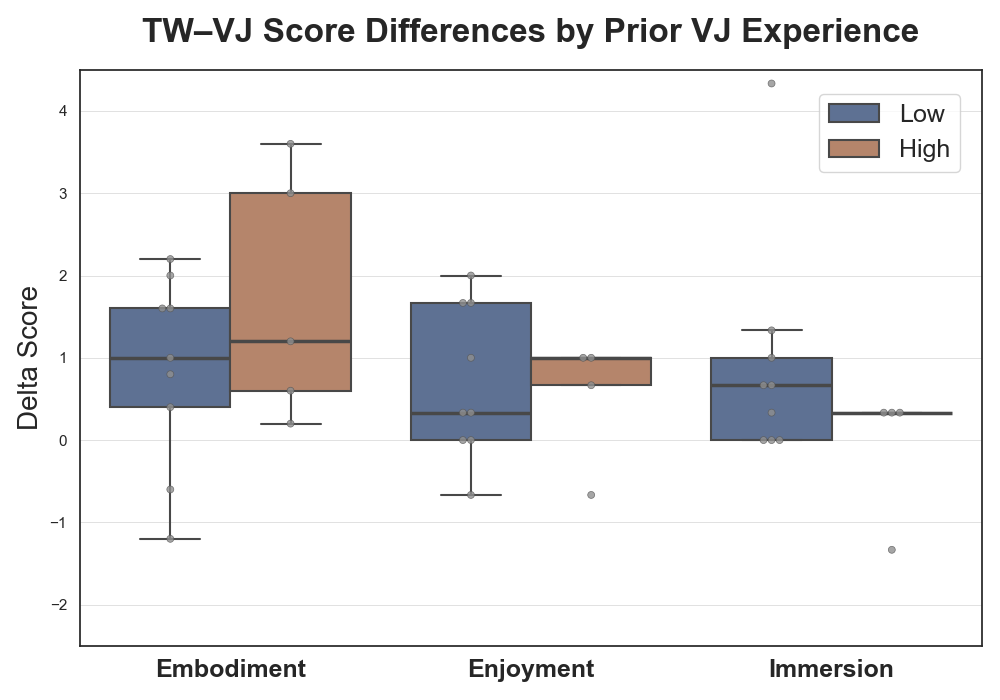}
  \caption{Box plots of TW–VJ score differences across joystick experience groups.  
No statistically significant differences were observed across sections.
"Low" and "High" indicate low and high VJ experience groups.}
  \label{fig:TW-VJ-VJgroup}
\end{figure}

To test whether prior joystick experience influenced participants’ relative preference for TW over VJ, we analyzed TW–VJ score differences using both correlation and group comparisons.
Although no statistically significant relationships were observed, all three sections showed negligible correlations with joystick experience:
Embodiment ($\rho = 0.108$, $p = 0.71$), Enjoyment ($\rho = 0.083$, $p = 0.78$), and Immersion ($\rho = 0.061$, $p = 0.84$),
suggesting that joystick familiarity had minimal impact on participants’ comparative ratings of the two interfaces.

To further assess potential group-level effects, we conducted Mann–Whitney U tests comparing low- and high-experience participants.
Participants were split into low (1–3) and high (4–5) experience groups based on their self-reported virtual joystick familiarity.
The tests revealed similarly non-significant results:
Embodiment ($U = 17.0$, $p = 0.505$, $r = 0.196$), Enjoyment ($U = 19.0$, $p = 0.687$, $r = 0.125$), and Immersion ($U = 25.5$, $p = 0.735$, $r = 0.107$). 
As shown in Figure~\ref{fig:TW-VJ-VJgroup}, participants with higher joystick experience reported slightly greater TW–VJ gains in Embodiment and Enjoyment,
whereas those with lower experience showed somewhat higher gains in Immersion. These differences, however, were not statistically significant, and all effect sizes were in the small range.

\textcolor{rv}{We also conducted additional TW-only analyses, which showed no significant effects of joystick experience (see Section~4 of the supplementary document).}
\textcolor{rv2}{In addition, analysis of participants’
selected touch area sizes revealed no significant differences in
embodiment, enjoyment, or immersion ratings across the chosen
sizes (see Section~8 of the supplementary document).}

{\color{rv}

\subsubsection{User Feedback (Summary)}

We summarize key insights from post-task interviews conducted to complement the quantitative findings. Participants broadly described TouchWalker as novel, engaging, and immersive, highlighting its embodied interaction style and tactile immediacy. Full interview responses are provided in Section~5 of the supplementary document.

\paragraph{Embodiment and Bodily Involvement}
Participants frequently reported a stronger sense of bodily connection compared to VJ, describing finger walking as rhythmically and spatially aligned with real-world walking. Many noted that the direct mapping between finger taps and avatar footsteps enhanced embodiment, although some also experienced dissonance due to using fingers instead of feet. Some users reported a sense of disconnection when the avatar did not respond as intended, which led to confusion or reduced control.

    \paragraph{Enjoyment and Novelty}
Enjoyment was often attributed to the novelty and expressiveness of the interaction, with several users describing the walking gesture itself as intrinsically fun. However, a few participants noted that unfamiliarity or input difficulty occasionally reduced enjoyment. Some also mentioned that the walking interaction itself became a source of enjoyment, rather than merely a way to reach a goal.

\paragraph{Immersion and Focused Engagement}
TW was seen as more immersive and cognitively engaging, with users attributing this to the need for continuous attentional focus and physical coordination. Several participants noted that directly controlling movement through finger gestures—rather than a joystick—created a stronger connection with the avatar, leading to a more focused and realistic experience.

\paragraph{Intuitiveness and Learnability}
Views on intuitiveness were mixed. While some participants found the walking metaphor natural, others reported initial confusion—especially with lateral movement. Several users said that control improved significantly after gaining familiarity with the rhythm and spatial mapping.
Some also noted that VJ input was more immediately graspable due to familiarity.

\paragraph{Control and Precision}
Participants appreciated TW’s precise control over foot placement, which was particularly useful in tasks requiring careful stepping. However, some found directional control more difficult to execute accurately, especially under time pressure. VJ was seen as more immediately controllable, but less expressive.

\paragraph{Perceived Use Cases, Physical Effort and Discomfort}
Many users suggested TW would be especially suitable for gameplay genres requiring deliberate movement, such as puzzle, stealth, or platformers. Finally, while fatigue was not widely reported, one participant mentioned mild discomfort from hand posture, indicating possible ergonomic considerations for extended use.
}

\subsection{Objective Evaluation}

\subsubsection{Task-Performance Metrics}

\begin{figure}
  \centering
  \includegraphics[width=\linewidth]{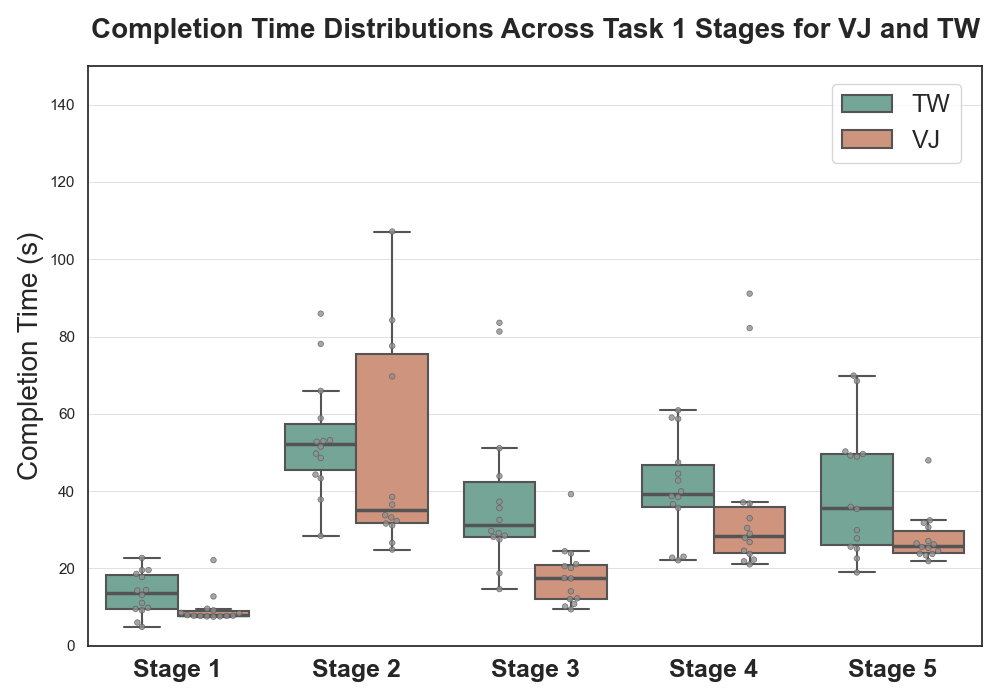}
  \caption{Box plots of completion times for each stage of Task 1, comparing TW and VJ input methods.
}
  \label{fig:T1-completion-time}
\end{figure}

In addition to subjective measures, we analyzed task performance as a secondary metric using data collected during the user study.

\paragraph{Stage Completion Time}
Figure~\ref{fig:T1-completion-time} illustrates completion time distributions for the Task 1 stages.
\textcolor{rv}{In Stage 1 (Hole Avoidance), participants completed the task significantly faster using VJ than with TW, as indicated by a paired t-test ($p=0.0187$, $d=0.718$).
For Stage 2 (Speed Control), no statistically significant difference was found between the two methods (Wilcoxon, $p=0.9032$); however, TW exhibited a noticeably narrower interquartile range, indicating more stable control under dynamic speed requirements.
In Stage 3 (Narrow Path), VJ again enabled significantly faster completion times (paired t-test, $p=0.0030$, $d=0.973$), with TW showing both longer times and greater variability, suggesting difficulty navigating fine spatial constraints.
Stage 4 (Foot Buttons) did not show a significant difference (Wilcoxon, $p=0.1531$), yet TW demonstrated more consistent performance overall, as indicated by a narrower interquartile range despite a few outliers.
In Stage 5 (Rolling Obstacles), VJ outperformed TW with significantly faster (paired t-test, $p=0.0124$, $d=0.775$), reinforcing the advantage of joystick input for fast-paced obstacle avoidance.
Collectively, these results confirm that while VJ generally offers speed advantages, especially in tasks requiring quick or spatially constrained movement, TW may facilitate more consistent execution in tasks that emphasize rhythm or precise foot interaction.}
Detailed summary statistics for each stage are provided in Section~6 of the supplementary document.

\begin{figure}
  \centering
  \includegraphics[width=\linewidth]{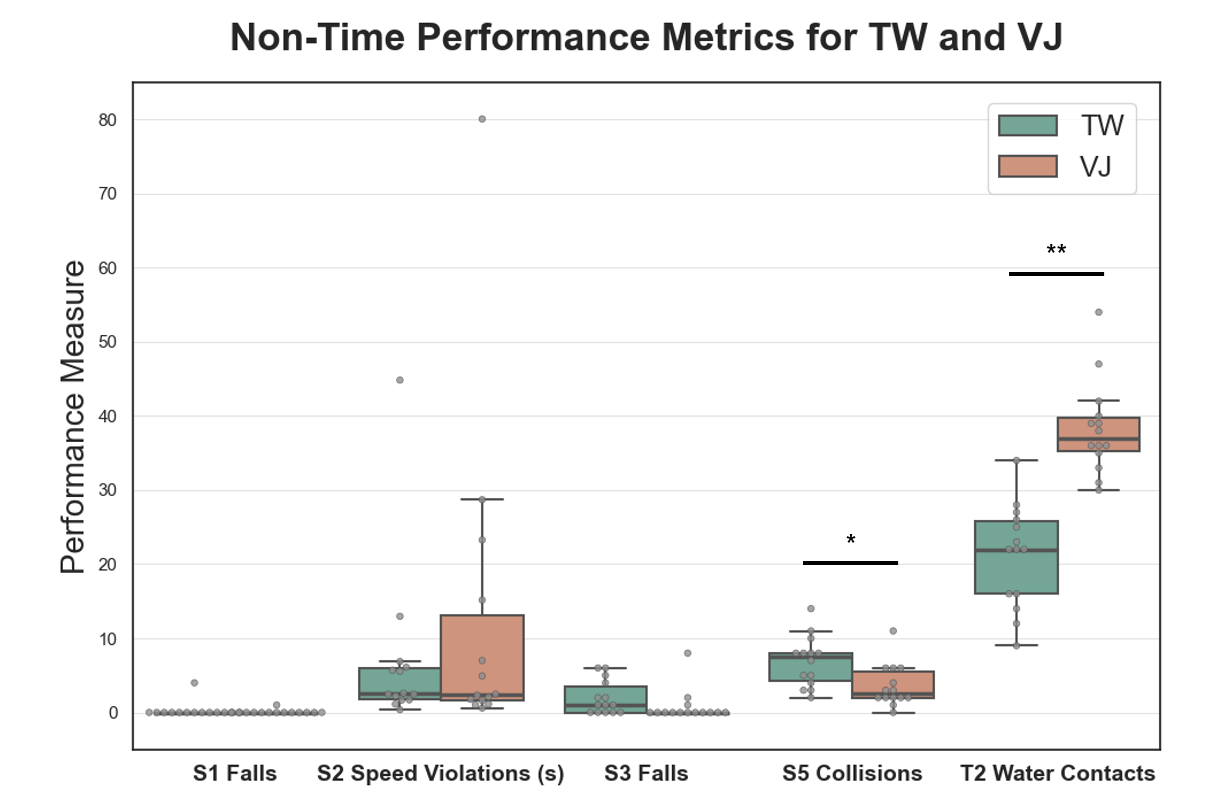}
  \caption{Boxplots of non-time performance metrics across TW and VJ input methods. Each pair compares performance on a specific task metric: number of falls (Stage 1, Stage 3), duration of speed violations (Stage 2), number of collisions (Stage 5), and number of water contacts (Task 2).}
  \label{fig:non-time-performance-metric}
\end{figure}

\paragraph{Non-time Performance Metrics}
Figure~\ref{fig:non-time-performance-metric} presents performance measures beyond completion time, offering further insight into the behavioral differences between TW and VJ.
\textcolor{rv}{In Stage 1, there was no significant difference in the number of falls between TW and VJ (Wilcoxon, $p=0.6547$), although TW had a higher average (M = 0.29) than VJ (M = 0.07). The medians were zero for both conditions, indicating that most participants completed the task without falling.
Stage 2 showed no significant difference in the duration of speed violations (Wilcoxon, $p=0.1961$), but TW again demonstrated more stable performance with less variability across participants.
For Stage 3, the difference in the number of falls approached significance (Wilcoxon, $p=0.0894$), with TW falling more often (M = 2.00) than VJ (M = 0.79), consistent with the increased completion time and variability.
In Stage 5, TW recorded a significantly higher number of collisions compared to VJ (Wilcoxon, $p=0.0116$, $r=0.721$), indicating a disadvantage for finger-walking in fast-paced obstacle avoidance.
Conversely, in Task 2 (Stepping Stones), TW had significantly fewer water contacts than VJ (Wilcoxon, $p=0.0001$, $r = 0.881$), reinforcing TW’s strength in tasks requiring precise and deliberate foot placement.}
Detailed summary statistics for each metric are provided in Section~7 of the supplementary document.

\textcolor{rv2}{We also examined whether participants’ preferred touch area size influenced objective performance metrics, but found no significant differences across size groups (see Section~8 of the supplementary document).}

\subsubsection{Ablation Study}

To evaluate each component of \textcolor{rv}{TouchWalker-MotionNet}, our motion generation model, we conducted an ablation study separate from the user study. All models were trained on subjects 1–4 from the LaFAN1 dataset~\cite{LaFAN1} and tested on subject 5. Input foot contact states and facing directions were derived from the test motion.

We compare \textcolor{rv}{the full TouchWalker-MotionNet} against several ablated variants using two proposed metrics and two standard ones:

\noindent \textbf{FCPE} (Foot Contact Position Error):  
Measures how closely the input contact position matches the contact foot position in the generated motion, computed as the average error per foot per frame.

\noindent \textbf{FCTE} (Foot Contact Timing Error):  
Measures the proportion of frames where the predicted contact status differs from the test motion, averaged across both feet. Contact status is determined by checking whether the predicted foot’s height and speed fall below the thresholds used to extract contact states from the test motion.

\noindent \textbf{MPJPE} (Mean Per Joint Position Error):  
Measures the average error in joint positions per frame, computed in the root frame.

\noindent \textbf{FS} (Foot Sliding):  
Measures the average horizontal movement of each foot during frames marked as contact in the test motion.

The ablated models used in the evaluation are as follows:

\noindent \textbf{w/o TransNet \& GRU}:  
A variant of \textcolor{rv}{TouchWalker-MotionNet} where PoseNet consists only of fully connected layers. This corresponds to the architecture proposed in \cite{MANN}, with input and output formats modified to match our setup.

\noindent \textbf{w/o TransNet}:  
A variant without TransNet, where only PoseNet with GRU is used.

\noindent \textbf{w/o GRU}:  
A variant where both TransNet and PoseNet use fully connected layers instead of GRUs.

\noindent \textbf{w/o $L_\mathrm{rec}$, w/o $L_\mathrm{FK}$, w/o $L_\mathrm{ct}$, w/o $L_\mathrm{ct\_trans}$}:  
Variants of our model trained without each respective loss term.

\begin{table}[h]
\center
    \caption{
Quantitative results from the ablation study.
}
 \scalebox{0.8}{
 \begin{tabular}{ l|c c c c }
 \toprule

                  & FCPE  $\downarrow$  & FCTE $\downarrow$   & MPJPE $\downarrow$  &  FS $\downarrow$           \\
                  & (cm/frame)        & (\%)             & (cm/frame)  &  (cm/frame)  \\   
                  \midrule

     w/o TransNet\&GRU         & 17.23  & 34.70  & 20.74  & 1.66  \\
     w/o TransNet     & 35.70  & 40.56  & 66.33  & 1.56  \\
     w/o GRU        & 24.25  & 38.55  & 31.29  & \textbf{1.40}  \\
     w/o $L_\mathrm{rec}$                & 12.20  & 33.00  & 25.92  & 2.50  \\
     w/o $L_\mathrm{FK}$                 & 37.02  & 48.12  & 122.89  & 3.40  \\
     w/o $L_\mathrm{ct}$            & 30.26  & 34.79  & 33.66  & 1.97  \\
     w/o $L_\mathrm{ct\_trans}$ & 10.13  & 26.06  & 20.19  & 1.68  \\
     Full model                                & \textbf{7.04}  & \textbf{24.42}  & \textbf{17.47}  & 1.68  \\
\bottomrule

\end{tabular}
}
\label{tbl:comparisons} 
\end{table}

As shown in Table~\ref{tbl:comparisons}, our full model outperforms all ablated variant in FCPE, FCTE, and MPJPE, demonstrating its ability to generate motions that align closely with user inputs and contextual cues. However, this responsiveness can cause rapid foot movement when input and avatar state diverge, leading to a relatively high FS.

\textbf{w/o TransNet \& GRU} exhibits minimal joint motion regardless of input, leading to forward translation without proper leg movement, which results in persistent foot sliding and high FCPE, FCTE, and MPJPE.
\textbf{w/o TransNet} shows the second-highest FCPE, FCTE, and MPJPE, reflecting its failure to produce appropriate root and posture dynamics—such as gradual body collapse or forward motion during backward walking.
\textbf{w/o GRU} shows poor overall motion quality, particularly in airborne actions such as running, due to its limited ability to leverage past context. Its low FS likely stems from overly conservative foot motion.
Removing $L_\mathrm{rec}$ leads to a nearly static avatar that fails to respond to input, resulting in high errors across all metrics.
Removing $L_\mathrm{FK}$ leads to severely degraded motion, with the avatar often collapsing or sinking below the ground, producing highest errors.  
In both cases, some slight alternation of the legs is observed, likely due to the remaining contact-related loss terms.
Excluding $L_\mathrm{ct}$ increases foot sliding and produces less contextually coherent motions, reflected in higher FCPE, FCTE, and MPJPE. Omitting $L_\mathrm{ct\_trans}$ has a milder effect on quantitative results, but can lead to occasional foot drift and instability.

\section{Analysis and Discussion}

\subsection{Subjective Experience and Benefits}

Our user study results confirm that TouchWalker (TW) outperformed the virtual joystick (VJ) across all subjective dimensions—embodiment, enjoyment, and immersion—supporting H1.
As shown in Figure 7, these differences were statistically significant, indicating a clear preference for TW.
We attribute this improvement to TW’s emphasis on foot-ground contact and tactile interaction, which foster a stronger sense of bodily involvement and contribute to more engaging and immersive user experiences.

\textcolor{rv}{We attribute this improvement not only to the embodied interaction style of finger-walking input but also to the underlying motion generation model. Unlike VJ, which replays predefined animation clips based on avatar speed and direction, TW uses a neural network (TouchWalker-MotionNet) to synthesize avatar motion per frame in real time from finger input. This allows the avatar’s foot placement and gait rhythm to closely follow user input, enabling expressive and responsive control. Such direct coupling between touch gestures and full-body motion reinforced the sensation of bodily involvement and enhanced user engagement.}

Qualitative feedback further supports these findings. Participants frequently described TW as more “bodily connected” and “engaging,” attributing this to the sensation of foot-ground contact and the tactile link between input and motion. Some also noted the rhythmic alternation of finger taps as contributing to a stronger sense of control and presence. While a few users experienced mild dissonance due to the use of fingers instead of feet, the embodied interaction overall appeared to enhance focus and deepen immersion.

\subsection{Consistency Across Joystick Experience Levels}

To test H2, we examined whether joystick familiarity affected the difference in ratings between TW and VJ. The results did not support the hypothesis, as no significant effects were found. Both novice and experienced users consistently preferred TW over VJ, suggesting that the relative advantage of TW does not depend on prior joystick experience.

\subsection{Task-Specific Strengths and Limitations}

Performance metrics revealed nuanced trade-offs between the two input methods. While VJ enabled faster task completion overall, TW demonstrated more consistent or accurate control in tasks requiring deliberate timing and spatial placement.
For example, in Stage 2 (Speed Control) and Stage 4 (Foot Buttons), TW resulted in more consistent performance across users, with narrower variability in task execution. In Task 2 (Stepping Stones), it further demonstrated an advantage in precise foot placement, leading to fewer water contacts. These findings highlight TW’s potential in tasks requiring stable control and deliberate interaction.

By contrast, TW showed limitations in tasks involving obstacle avoidance or spatially constrained movement (e.g., Stage 3 and Stage 5). These tasks often require either quick reactions (e.g., obstacle avoidance in Stage 5) or precise spatial navigation (e.g., narrow path traversal in Stage 3), which may be less compatible with the physical and cognitive characteristics of finger-walking input.
Additionally, instead of rotating the avatar directly, our system inputs the desired facing direction into the motion generator to ensure natural and coherent movement. While this improves realism, it can introduce slight turning delays, potentially affecting responsiveness in time-critical tasks.

\textcolor{rv}{
The performance differences between TW and VJ stem not only from input modality but also from their motion generation mechanisms. While TW generates motion per frame from finger input, VJ replays predefined animation clips based on avatar speed and direction. As a result, VJ cannot reflect fine-grained differences in input timing or placement—critical for precision tasks like Task 2. In contrast, TW maps each touch to foot-ground contact in real time, enabling spatially precise control.
}

Overall, while the finger-walking approach supports expressive and immersive avatar control, it may not be optimal for high-speed or densely structured action sequences.

\subsection{Future Potential}

Taken together, our results suggest that TW offers a low-barrier, high-engagement locomotion interface for touchscreen devices. While it may require initial adaptation, its rhythmic and embodied interaction style offers strong subjective appeal and performs well in tasks involving precise control. Given these strengths, TW holds promise for integration into game genres emphasizing deliberate movement and immersion, such as puzzle, adventure, or stealth games. 

At the same time, user feedback highlighted areas for improvement. Directional control—particularly lateral movement—was often described as challenging under time pressure. And while TW’s fine-grained foot placement was appreciated in tasks demanding careful control, it also introduced greater cognitive and physical effort. Enhancing directional responsiveness and easing the input burden could help expand its applicability to faster-paced or more dynamic interactions.
\textcolor{rv}{In addition, one participant noted minor hand discomfort during finger walking, suggesting that prolonged use may introduce ergonomic challenges. Future work could explore interface refinements to reduce physical strain and support longer gameplay sessions.}

\section{Conclusion}

\textcolor{rv}{We introduced TouchWalker, a real-time touchscreen locomotion system comprising TouchWalker-MotionNet, a neural motion generation model, and TouchWalker-UI, a finger-walking interface.}
Through a user study comparing it with a virtual joystick, we found that TouchWalker significantly improved users' sense of embodiment, enjoyment, and immersion, regardless of prior joystick experience. While it showed advantages in tasks requiring precision and deliberate foot placement, challenges remained in spatially-constrained or rapid-response scenarios. These findings suggest that TouchWalker offers a promising alternative for expressive avatar control, particularly in applications emphasizing immersion and bodily involvement.
\textcolor{rv}{By enabling per-frame interpretation of rhythmic finger gestures, TouchWalker also reimagines the interaction paradigm itself—shifting finger-walking from a symbolic cue to a continuous control signal that can drive nuanced full-body motion.}

\acknowledgments{%
This work was supported by the National Research Foundation
of Korea (NRF) grant (RS-2023-00222776); and by Culture, Sports and
Tourism R\&D Program through the Korea Creative Content Agency
grant funded by the Ministry of Culture, Sports and Tourism in 2024
(RS-2024-00399136).
}

\bibliographystyle{abbrv-doi-hyperref}

\bibliography{ref}

@inproceedings{
tevet2023human,
title={Human Motion Diffusion Model},
author={Guy Tevet and Sigal Raab and Brian Gordon and Yoni Shafir and Daniel Cohen-or and Amit Haim Bermano},
booktitle={The Eleventh International Conference on Learning Representations },
year={2023},
}

@inproceedings{
shafir2024human,
title={Human Motion Diffusion as a Generative Prior},
author={Yoni Shafir and Guy Tevet and Roy Kapon and Amit Haim Bermano},
booktitle={The Twelfth International Conference on Learning Representations},
year={2024},
url={https://openreview.net/forum?id=dTpbEdN9kr}
}

@inproceedings{cohan_flexible_2024,
	address = {New York, NY, USA},
	series = {{SIGGRAPH} '24},
	title = {Flexible {Motion} {In}-betweening with {Diffusion} {Models}},
	isbn = {9798400705250},
	booktitle = {{ACM} {SIGGRAPH} 2024 {Conference} {Papers}},
	publisher = {Association for Computing Machinery},
	author = {Cohan, Setareh and Tevet, Guy and Reda, Daniele and Peng, Xue Bin and van de Panne, Michiel},
	month = jul,
	year = {2024},
	pages = {1--9},
}

@inproceedings{tang_rsmt_2023,
	series = {{SIGGRAPH} '23},
	title = {{RSMT}: {Real}-time {Stylized} {Motion} {Transition} for {Characters}},
	shorttitle = {{RSMT}},
	booktitle = {{ACM} {SIGGRAPH} 2023 {Conference} {Proceedings}},
	publisher = {Association for Computing Machinery},
	author = {Tang, Xiangjun and Wu, Linjun and Wang, He and Hu, Bo and Gong, Xu and Liao, Yuchen and Li, Songnan and Kou, Qilong and Jin, Xiaogang},
	year = {2023},
	pages = {1--10},
}

@article{sun_augmented_2022,
	title = {Augmented tactile-perception and haptic-feedback rings as human-machine interfaces aiming for immersive interactions},
	volume = {13},
	copyright = {2022 The Author(s)},
	issn = {2041-1723},
	language = {en},
	number = {1},
	urldate = {2025-04-11},
	journal = {Nature Communications},
	author = {Sun, Zhongda and Zhu, Minglu and Shan, Xuechuan and Lee, Chengkuo},
	month = sep,
	year = {2022},
	note = {Publisher: Nature Publishing Group},
	pages = {5224},
}

@article{li_haptics-mediated_2024,
	title = {Haptics-mediated virtual embodiment: {Impact} of a wearable avatar-controlling system with kinesthetic gloves on embodiment in {VR}},
	volume = {5},
	issn = {2673-4192},
	shorttitle = {Haptics-mediated virtual embodiment},
	language = {English},
	journal = {Frontiers in Virtual Reality},
	author = {Li, Zhenxing and Bujic, Mila and Buruk, Oguz "Oz" and Bampouni, Elpida and Jarvela, Simo and Hamari, Juho},
	month = sep,
	year = {2024},
	note = {Publisher: Frontiers},
}

@article{kilteni_sense_2012,
	title = {The {Sense} of {Embodiment} in {Virtual} {Reality}},
	volume = {21},
	issn = {1054-7460},
	number = {4},
	urldate = {2025-04-08},
	journal = {Presence},
	author = {Kilteni, Konstantina and Groten, Raphaela and Slater, Mel},
	month = jan,
	year = {2012},
	pages = {373--387},
}

@book{dourish_where_2001,
	title = {Where the {Action} {Is}: {The} {Foundations} of {Embodied} {Interaction}},
	isbn = {978-0-262-25605-6},
	shorttitle = {Where the {Action} {Is}},
	language = {en},
	urldate = {2025-04-08},
	publisher = {The MIT Press},
	author = {Dourish, Paul},
	month = sep,
	year = {2001},
}

@inproceedings{hung_fingerpuppet_2024,
	address = {New York, NY, USA},
	series = {{CHI} {EA} '24},
	title = {{FingerPuppet}: {Finger}-{Walking} {Performance}-based {Puppetry} for {Human} {Avatar}},
	isbn = {9798400703317},
	shorttitle = {{FingerPuppet}},
	booktitle = {Extended {Abstracts} of the 2024 {CHI} {Conference} on {Human} {Factors} in {Computing} {Systems}},
	publisher = {Association for Computing Machinery},
	author = {Hung, Ching-Wen and Liang, Chung-Han and Chen, Bing-Yu},
	month = may,
	year = {2024},
	pages = {1--6},
}

@inproceedings{hung_puppeteer_2022,
	address = {New York, NY, USA},
	series = {{VRST} '22},
	title = {Puppeteer: {Exploring} {Intuitive} {Hand} {Gestures} and {Upper}-{Body} {Postures} for {Manipulating} {Human} {Avatar} {Actions}},
	isbn = {978-1-4503-9889-3},
	shorttitle = {Puppeteer},
	booktitle = {Proceedings of the 28th {ACM} {Symposium} on {Virtual} {Reality} {Software} and {Technology}},
	publisher = {Association for Computing Machinery},
	author = {Hung, Ching-Wen and Chang, Ruei-Che and Chen, Hong-Sheng and Liang, Chung Han and Chan, Liwei and Chen, Bing-Yu},
	month = nov,
	year = {2022},
	pages = {1--11},
}

@inproceedings{zaman_touchscreens_2010,
	address = {New York, NY, USA},
	series = {Futureplay '10},
	title = {Touchscreens vs. traditional controllers in handheld gaming},
	isbn = {978-1-4503-0235-7},
	booktitle = {Proceedings of the {International} {Academic} {Conference} on the {Future} of {Game} {Design} and {Technology}},
	publisher = {Association for Computing Machinery},
	author = {Zaman, Loutfouz and Natapov, Daniel and Teather, Robert J.},
	month = may,
	year = {2010},
	pages = {183--190},
}

@inproceedings{rigby_development_2019,
	address = {New York, NY, USA},
	series = {{TVX} '19},
	title = {Development of a {Questionnaire} to {Measure} {Immersion} in {Video} {Media}: {The} {Film} {IEQ}},
	isbn = {978-1-4503-6017-3},
	shorttitle = {Development of a {Questionnaire} to {Measure} {Immersion} in {Video} {Media}},
	booktitle = {Proceedings of the 2019 {ACM} {International} {Conference} on {Interactive} {Experiences} for {TV} and {Online} {Video}},
	publisher = {Association for Computing Machinery},
	author = {Rigby, Jacob M. and Brumby, Duncan P and Gould, Sandy J. J. and Cox, Anna L},
	month = jun,
	year = {2019},
	pages = {35--46},
}

@misc{IMI,
  author       = {{Self-Determination Theory}},
  title        = {Intrinsic Motivation Inventory (IMI)},
  year         = {n.d.},
  howpublished = {\url{https://selfdeterminationtheory.org/wp-content/uploads/2022/02/IMI_Complete.pdf}},
  note         = {Accessed: 2025-04-02}
}

@article{jennett_measuring_2008,
	title = {Measuring and defining the experience of immersion in games},
	volume = {66},
	issn = {1071-5819},
	number = {9},
	urldate = {2025-03-26},
	journal = {International Journal of Human-Computer Studies},
	author = {Jennett, Charlene and Cox, Anna L. and Cairns, Paul and Dhoparee, Samira and Epps, Andrew and Tijs, Tim and Walton, Alison},
	month = sep,
	year = {2008},
	pages = {641--661},
}

@article{ryan_self-determination_2000,
	title = {Self-determination theory and the facilitation of intrinsic motivation, social development, and well-being},
	volume = {55},
	issn = {1935-990X},
	number = {1},
	journal = {American Psychologist},
	author = {Ryan, Richard M. and Deci, Edward L.},
	year = {2000},
	note = {Place: US
Publisher: American Psychological Association},
	pages = {68--78},
}

@article{peck_avatar_2021,
	title = {Avatar {Embodiment}. {A} {Standardized} {Questionnaire}},
	volume = {1},
	issn = {2673-4192},
	language = {English},
	urldate = {2024-08-23},
	journal = {Frontiers in Virtual Reality},
	author = {Peck, Tabitha C. and Gonzalez-Franco, Mar},
	month = feb,
	year = {2021},
	note = {Publisher: Frontiers},
}

@article{guiard_asymmetric_1987,
	title = {Asymmetric division of labor in human skilled bimanual action: the kinematic chain as a model},
	volume = {19},
	issn = {0022-2895},
	shorttitle = {Asymmetric division of labor in human skilled bimanual action},
	language = {eng},
	number = {4},
	journal = {Journal of Motor Behavior},
	author = {Guiard, Y.},
	month = dec,
	year = {1987},
	pmid = {15136274},
	pages = {486--517},
}

@article{teather_tilt-touch_2017,
	title = {Tilt-{Touch} synergy: {Input} control for “dual-analog” style mobile games},
	volume = {21},
	issn = {1875-9521},
	shorttitle = {Tilt-{Touch} synergy},
	urldate = {2025-03-28},
	journal = {Entertainment Computing},
	author = {Teather, Robert J. and Roth, Andrew and MacKenzie, I. Scott},
	month = jun,
	year = {2017},
	pages = {33--43},
}

@inproceedings{baxter_virtual_2023,
	address = {New York, NY, USA},
	series = {{MIG} '23},
	title = {Virtual {Joystick} {Control} {Sensitivity} and {Usage} {Patterns} in a {Large}-{Scale} {Touchscreen}-{Based} {Mobile} {Game} {Study}},
	isbn = {9798400703935},
	booktitle = {Proceedings of the 16th {ACM} {SIGGRAPH} {Conference} on {Motion}, {Interaction} and {Games}},
	publisher = {Association for Computing Machinery},
	author = {Baxter, John and Adamson, Torin and Hasan, Yazied and Yousefi, Mohammad and Obregon, Lidia and Carter, Evan and Tapia, Lydia},
	month = nov,
	year = {2023},
	pages = {1--6},
}

@inproceedings{lee_novel_2024,
	title = {A {Novel} {Approach} for {Virtual} {Locomotion} {Gesture} {Classification}: {Self}-{Teaching} {Vision} {Transformer} for a {Carpet}-{Type} {Tactile} {Sensor}},
	shorttitle = {A {Novel} {Approach} for {Virtual} {Locomotion} {Gesture} {Classification}},
	booktitle = {2024 {IEEE} {Conference} {Virtual} {Reality} and {3D} {User} {Interfaces} ({VR})},
	author = {Lee, Sung-Ha and Joo, Ho-Taek and Chung, Insik and Park, Donghyeok and Choi, Yunho and Kim, Kyung-Joong},
	month = mar,
	year = {2024},
	note = {ISSN: 2642-5254},
	pages = {461--471},
}

@article{bergamin_drecon_2019,
	title = {{DReCon}: data-driven responsive control of physics-based characters},
	volume = {38},
	issn = {0730-0301},
	shorttitle = {{DReCon}},
	number = {6},
	urldate = {2020-02-21},
	journal = {ACM Transactions on Graphics (TOG)},
	author = {Bergamin, Kevin and Clavet, Simon and Holden, Daniel and Forbes, James Richard},
	month = nov,
	year = {2019},
}

@article{starke_neural_2021,
	title = {Neural animation layering for synthesizing martial arts movements},
	volume = {40},
	issn = {0730-0301},
	number = {4},
	urldate = {2022-04-24},
	journal = {ACM Transactions on Graphics},
	author = {Starke, Sebastian and Zhao, Yiwei and Zinno, Fabio and Komura, Taku},
	year = {2021},
}

@article{holden_phase-functioned_2017,
	title = {Phase-functioned {Neural} {Networks} for {Character} {Control}},
	volume = {36},
	issn = {0730-0301},
	number = {4},
	urldate = {2017-12-21},
	journal = {ACM Transactions on Graphics},
	author = {Holden, Daniel and Komura, Taku and Saito, Jun},
	month = jul,
	year = {2017},
}

@article{ling_character_2020,
	title = {Character controllers using motion {VAEs}},
	volume = {39},
	issn = {0730-0301},
	number = {4},
	journal = {ACM Transactions on Graphics},
	author = {Ling, Hung Yu and Zinno, Fabio and Cheng, George and Van De Panne, Michiel},
	year = {2020},
}

@article{MANN,
  title={Mode-adaptive neural networks for quadruped motion control},
  author={He Zhang and Sebastian Starke and Taku Komura and Jun Saito},
  journal={ACM Transactions on Graphics},
  year={2018},
  volume={37},
}

@inproceedings{NeMF,
    author = {He, Chengan and Saito, Jun and Zachary, James and Rushmeier, Holly and Zhou, Yi},
    title = {NeMF: Neural Motion Fields for Kinematic Animation},
    booktitle = {NeurIPS},
 volume = {35},
    year = {2022}
}

@article{FootContactLoss,
author = {Lee, Kyungho and Lee, Seyoung and Lee, Jehee},
title = {Interactive Character Animation by Learning Multi-Objective Control},
year = {2018},
volume = {37},
journal = {ACM Transactions on Graphics},
}

@inproceedings{lockwood_finger_2012,
	address = {Goslar, DEU},
	series = {{SCA} '12},
	title = {Finger walking: motion editing with contact-based hand performance},
	isbn = {978-3-905674-37-8},
	shorttitle = {Finger walking},
	urldate = {2024-08-06},
	booktitle = {Proceedings of the {ACM} {SIGGRAPH}/{Eurographics} {Symposium} on {Computer} {Animation}},
	publisher = {Eurographics Association},
	author = {Lockwood, Noah and Singh, Karan},
	month = jul,
	year = {2012},
	pages = {43--52},
}

@InProceedings{FKLayer,
  author = {Villegas, Ruben and Yang, Jimei and Ceylan, Duygu and Lee, Honglak},
  title = {Neural Kinematic Networks for Unsupervised Motion Retargetting},
  booktitle = {The IEEE Conference on Computer Vision and Pattern Recognition (CVPR)},
  year = {2018}
}

@article{LaFAN1,
author = {Harvey, F\'{e}lix G. and Yurick, Mike and Nowrouzezahrai, Derek and Pal, Christopher},
title = {Robust Motion In-Betweening},
year = {2020},
volume = {39},
journal = {ACM Transactions on Graphics},
}

@article{ARAnimator,
author = {Ye, Hui and Kwan, Kin Chung and Su, Wanchao and Fu, Hongbo},
title = {ARAnimator: in-situ character animation in mobile AR with user-defined motion gestures},
year = {2020},
issue_date = {August 2020},
publisher = {Association for Computing Machinery},
volume = {39},
number = {4},
issn = {0730-0301},
journal = {ACM Trans. Graph.},
month = aug,
articleno = {83},
numpages = {12},
}

@inproceedings{DoubleDoodles,
author = {Chen, Ruizhao and Pan, Ye and Deng, Zhigang and Wang, Lili and Ma, Lizhuang},
title = {Double Doodles: Sketching Animation in Immersive Environment With 3+6 DOFs Motion Gestures},
year = {2023},
isbn = {9798400701085},
publisher = {Association for Computing Machinery},
address = {New York, NY, USA},
booktitle = {Proceedings of the 31st ACM International Conference on Multimedia},
pages = {6998–7006},
numpages = {9},
series = {MM '23}
}

@inproceedings{HandAvatar,
author = {Jiang, Yu and Li, Zhipeng and He, Mufei and Lindlbauer, David and Yan, Yukang},
title = {HandAvatar: Embodying Non-Humanoid Virtual Avatars through Hands},
year = {2023},
isbn = {9781450394215},
publisher = {Association for Computing Machinery},
booktitle = {Proceedings of the 2023 CHI Conference on Human Factors in Computing Systems},
articleno = {309},
numpages = {17},
series = {CHI '23}
}

@INPROCEEDINGS{FootPad,
  author={Lee, Sung-Ha and Joo, Ho-Taek and Chung, Insik and Park, Donghyeok and Choi, Yunho and Kim, Kyung-Joong},
  booktitle={2023 IEEE International Symposium on Mixed and Augmented Reality Adjunct (ISMAR-Adjunct)}, 
  title={A Novel Approach for Virtual Locomotion Gesture Classification: Self-Teaching Vision Transformer for a Carpet-Type Tactile Sensor}, 
  year={2023},
  volume={},
  number={},
  pages={369-370},
}

@inproceedings{FWIP,
author = {Kim, Ji-Sun and Gra\v{c}anin, Denis and Matkovi\'{c}, Kre\v{s}imir and Quek, Francis},
title = {Finger Walking in Place (FWIP): A Traveling Technique in Virtual Environments},
year = {2008},
isbn = {9783540854104},
publisher = {Springer-Verlag},
booktitle = {Proceedings of the 9th International Symposium on Smart Graphics},
pages = {58–69},
numpages = {12},
series = {SG '08}
}

\end{document}

\firstsection{Training Details}
\maketitle

\textcolor{rv}{TouchWalker-MotionNet} has been trained and evaluated using motion data consisting of various locomotion behaviors from the LaFAN1 dataset~\cite{LaFAN1}.
Detailed information can be found in Table~\ref{tbl:dataset_details}.

\begin{table}[h]
    \center
    \caption{Dataset details.
    }
    \scalebox{0.9}{
        \begin{tabular}{ c|c c c}\hline

            LaFAN1 theme & \# of sequences & \# of frames  &  seconds   \\
            \hline

            Walk         & 12 & 86871 & 2896  \\
            \hline

            Run          & 4 & 28960  & 965    \\
            \hline

            Jumps        & 3 & 22022  & 734    \\
            \hline

            Sprint       & 2 & 16388  & 546   \\
            \hline
	  
            Aiming       & 5 & 41884  & 1396   \\
            \hline

        \end{tabular}
        }
        \label{tbl:dataset_details} 
\end{table}

We trained the model using the AdamW optimizer with an initial learning rate of $1.0 \times 10^{-4}$ and a weight decay of $2.5 \times 10^{-3}$ on Pytorch 2.3.1 with cuda 11.8, running on Python 3.10.13. Training was performed with a batch size of 256 using random batch sampling, and took approximately 3 hours on an Nvidia GeForce RTX 3090 GPU.

\section{Runtime Implementation Details}

The trained network was exported to the ONNX format and integrated into a Unity-based client application using the ONNX Runtime plugin. We used Unity Editor version 2023.2.20f1 to build the user study applications, targeting Android OS version 13 (Tiramisu). The motion generation module runs fully on-device at inference time, taking input from the touchscreen and producing avatar poses frame-by-frame. All experiments were conducted on a Galaxy Tab S7+ (12.4-inch display), where the system maintained consistently high performance. The average frame rate remained close to 30 FPS across all tasks: for Task 1, Stages 1 through 5, the FPS was 29.41, 29.91, 29.87, 29.86, and 29.86, respectively; for Task 2, the average FPS was 29.59.

To reduce potential fluctuations from abrupt input changes, we apply exponential smoothing as in \cite{bergamin_drecon_2019} with a smoothing factor $\beta = 0.7$: $\tilde{\mathbf c}_{t+1} \leftarrow \beta \tilde{\mathbf c}_{t+1} + (1 - \beta) \tilde{\mathbf c}_t.$
The smoothed output is used to render the avatar in the next frame and to update the input history for subsequent predictions.

To support locomotion on uneven terrain, we apply inverse kinematics (IK) to refine the avatar’s lower body after motion generation. The horizontal positions of the root and feet remain fixed, while their vertical positions are adjusted using terrain data. Each foot’s vertical position is computed by adding the predicted foot height to the terrain height at its horizontal location.
If at least one foot is in contact, the root height is set based on the height of the contacting foot (or the lower one if both are in contact). When both feet are airborne, the root height is gradually lowered from the height based on the last contacted foot to the height based on the terrain. This prevents sudden drops when the root hovers over low terrain (e.g., water) while the avatar appears to stand on a higher platform.
Finally, analytic two-joint IK is applied to align the legs with the updated foot positions.

\section{Questionnaire for User Study}

To evaluate the user experience of TouchWalker across various aspects such as embodiment, enjoyment, and immersion, we constructed a 
11-item
questionnaire by selecting relevant items from several original instruments.

\begin{description}

\item[Embodiment]
To assess embodiment, we selected five items from the Embodiment Questionnaire (EQ) \cite{peck_avatar_2021}. The numbers in parentheses indicate the corresponding item numbers from the revised version of the questionnaire presented in \cite{peck_avatar_2021}. The original body-related phrasing of each item was adapted to suit the context of our experiment, following the guidelines provided in the original paper.
\begin{itemize}
    \item I felt as if the character’s legs were my body part. (R10)
    \item I felt as if the movements of the character were influencing my own movements. (R3)
    \item I felt as if I was located where I saw the character. (R12)
    \item It felt like I could control the character as if it was my own body. (R13)
    \item It seemed as if the touch I felt was caused by the fingers touching the virtual ground. (R15)
\end{itemize}

\item[Enjoyment]
Enjoyment was assessed using the Intrinsic Motivation Inventory (IMI) \cite{ryan_self-determination_2000}, specifically through three items from the Interest/Enjoyment section of \cite{IMI}. To more accurately capture the enjoyment associated with each control method, we replaced the phrase "this activity" in the original items with "this control" in our adaptation.
\begin{itemize}
    \item I enjoyed doing this control very much.
    \item This control was fun to do.
    \item I would describe this control as very interesting.
\end{itemize}

\item[Immersion]
Immersion was assessed using the Immersive Experience Questionnaire (IEQ) \cite{jennett_measuring_2008}, specifically through three items referenced in \cite{rigby_development_2019} from the original question set. The items were rephrased from an interrogative form to a declarative statement to better match the Likert-scale response format used in our questionnaire. The numbers in parentheses indicate the original item numbers as referenced in \cite{rigby_development_2019}.
\begin{itemize}
    \item I felt that I was focused on the game. (2)
    \item I put a lot of effort into playing the game. (3)
    \item I felt that I was trying my best. (4)
\end{itemize}

\end{description}

{\color{rv}
\section{Additional Analysis: TW-only Ratings and Joystick Familiarity}

As a complementary analysis, we also examined whether joystick experience influenced participants’ evaluation of the TW interface alone.
Spearman correlations between joystick experience and TW-only scores again revealed no significant relationships:
Embodiment ($\rho = 0.077$, $p = 0.79$), Enjoyment ($\rho = -0.049$, $p = 0.87$), and Immersion ($\rho = 0.001$, $p = 0.99$).
Group-level comparisons also yielded non-significant results:  
Embodiment ($U = 15.0$, $p = 0.349$, $r = 0.267$),  
Enjoyment ($U = 20.0$, $p = 0.785$, $r = 0.089$), and  
Immersion ($U = 23.5$, $p = 0.946$, $r = 0.036$).  
The score distributions were largely overlapping, indicating that TW was evaluated consistently regardless of prior joystick familiarity.  
All effect sizes were small, further suggesting minimal influence of joystick experience on participants’ evaluations of the TW interface.

}

\section{Full User Feedback}

To complement the quantitative ratings, we conducted semi-structured interviews to better understand how participants perceived the TouchWalker interface. While the questionnaire focused on embodiment, enjoyment, and immersion, participants also spontaneously commented on other aspects such as intuitiveness, control precision, and potential use cases. These qualitative insights help contextualize and expand upon the survey results.

\paragraph{Embodiment and Bodily Involvement}

Many participants reported a stronger bodily connection with the avatar when using TW. They described the finger-walking gesture as “making it feel like my legs were actually moving” (P1), or that they “felt more like controlling the character’s body directly” (P3, P6, P8). Several noted that directly tapping the screen with alternating fingers resembled the rhythm and mechanics of real walking, which “made the avatar feel more like an extension of their own body” (P13).
In addition to movement style, several participants associated the sensation of foot-ground contact with a stronger sense of bodily involvement. As P8 explained, “Since the character’s foot touched exactly where my finger landed, it felt like it was me—or more precisely, my finger—touching the ground,” highlighting tactile overlap between the user and the avatar. P4 similarly noted that finger walking “somewhat felt like contact with the surface,” despite also commenting that “There was some dissonance, since it’s my fingers doing the walking, not my feet,” reflecting a mix of grounded sensation and bodily dissonance.
Other participants also noted a sense of disconnection when the avatar did not move as intended, contributing to a feeling of confusion or reduced control (P9).

\paragraph{Enjoyment and Novelty}

TW was frequently described as “fun because it felt new and different” (P6). For many, the novelty itself contributed to enjoyment: “It’s a new way to interact, and that alone made it more interesting” (P11). Several participants emphasized that the act of walking became an engaging activity in itself, rather than a means to an end (P3, P13). P3 shared, “The finger walking interaction itself became the goal—it wasn't just about reaching a destination, but about enjoying the control.”
A few, however, noted that the enjoyment was hindered at times by control difficulty or learning effort (P10). P10 remarked, “Since it was my first time trying this control method, it felt unfamiliar and somewhat difficult to adapt to, so I’m not sure yet if I fully enjoyed it.”

\paragraph{Immersion and Focused Engagement}

Several participants described a heightened sense of immersion when using TW, often tied to the increased attentional demand and physical involvement. Some noted that coordinating finger movements for walking required more focus, which in turn deepened their engagement with the task (P8, P12). Others mentioned that controlling movement through their own fingers, rather than a joystick, created a stronger sense of connection with the avatar’s motion (P1, P6, P10).
As P1 explained, “When I ran with my fingers, the character ran too, and when I walked, the character walked—it made me feel more focused in the game.” Similarly, P8 noted, “I had to concentrate more on moving my fingers, which actually made me feel more immersed.”
This embodied interaction was seen as contributing not only to realism but also to a more focused and immersive experience overall.

\paragraph{Intuitiveness and Learnability}

Participants expressed mixed impressions regarding the intuitiveness of TW. Some highlighted its natural mapping to walking behavior (P2, P4, P8), while others found movement direction control (especially lateral movement) initially confusing (P1, P5, P7). Several commented that once the rhythm and mechanism were learned, actions became more fluid and intuitive (P3, P12). Nonetheless, joystick input was still perceived by some as more immediately graspable due to familiarity or simpler directional mapping (P7, P10).
P2 shared, “It felt natural. Like tapping to walk—my brain just got it.” In contrast, P5 noted, “At first, I was confused about which direction to go. I had to get used to the rhythm.”

\paragraph{Control and Precision}

Participants demonstrated a nuanced understanding of TW’s control dynamics. Some appreciated the ability to determine exactly where each footstep would land, particularly in tasks that required careful placement (P4, P10). As P10 noted, “When crossing the river, with finger walking, you can directly step on each stone. Although I stepped in the water sometimes because I wasn’t used to it yet, I think once familiar, this would feel much more intuitive.”
However, several participants mentioned specific challenges affecting precision. Directional control, particularly lateral movement, was cited as confusing or difficult to execute accurately, at least initially (P1, P7, P9). P1 remarked, “Going forward and backward was easy, but moving left and right was confusing, especially under time pressure—I often went in the wrong direction.” 
In contrast, joystick input was often described as simpler to execute accurately from the beginning, but participants acknowledged it provided less nuanced or expressive physical control over avatar movement (P3, P6).

\paragraph{Perceived Use Cases}

Several participants speculated on possible applications of TW beyond the current study. Many felt that the interface could be particularly effective in game genres requiring cautious, deliberate movement—such as horror, puzzle, stealth, or platforming games (P6, P10, P13). For example, P13 remarked, “It would be really fun for games similar to Human: Fall Flat, where you step on switches to open doors and things like that.”
Others noted that while joystick input might remain preferable for fast-paced or action-heavy games, TW could offer a compelling alternative for slower or more immersive gameplay styles (P5, P11). P4 pointed out that the embodied input of TW could enhance tension in games requiring quick physical responses: “With joysticks, tension comes from simply pushing the stick harder to move forward faster, but with this, you physically need to move quickly with your fingers, so I think it could be useful in those situations.”
P11 added, “It’s a new kind of experience that people haven’t really tried before, so I think it could attract attention—especially if the game emphasizes careful movement or character control.”

{\color{rv}
\paragraph{Physical Effort and Discomfort}

While fatigue was not explicitly mentioned by participants during interviews, one participant (P11) noted some physical discomfort, stating: “I have to bend my hand to perform finger walking, so it can feel a bit uncomfortable.” This suggests that specific hand postures required for finger walking may contribute to physical strain. While such discomfort was not widely reported, it indicates that some users may experience mild ergonomic challenges, even if fatigue was not a prominent concern in our sample.
}

\section{Task 1 Stage Completion Time Statistics}

\begin{table}[ht]
\centering
\caption{Completion time statistics for each stage of Task 1 under TouchWalker (TW) and virtual joystick (VJ) input. All values are in seconds.}
\label{tab:task1-stage-times}
\begin{tabular}{llrrrr}
\toprule
\textbf{Stage} & \textbf{Method} & \textbf{Mean} & \textbf{Median} & \textbf{SD} & \textbf{IQR} \\
\midrule
\multirow{2}{*}{Stage 1} & TW & 13.63 & 13.74 & 5.47 & 8.80 \\
                         & VJ &  9.45 &  7.87 & 3.91 & 1.28 \\
\multirow{2}{*}{Stage 2} & TW & 53.68 & 52.17 & 15.05 & 12.06 \\
                         & VJ & 59.15 & 35.15 & 48.22 & 43.81 \\
\multirow{2}{*}{Stage 3} & TW & 38.71 & 31.13 & 20.65 & 14.00 \\
                         & VJ & 18.09 & 17.48 &  7.93 & 8.88 \\
\multirow{2}{*}{Stage 4} & TW & 40.78 & 39.35 & 12.89 & 10.79 \\
                         & VJ & 36.29 & 28.47 & 22.01 & 11.90 \\
\multirow{2}{*}{Stage 5} & TW & 39.85 & 35.67 & 16.48 & 23.36 \\
                         & VJ & 27.93 & 25.90 & 6.60  & 5.78 \\
\bottomrule
\end{tabular}
\end{table}

Table~\ref{tab:task1-stage-times} presents descriptive statistics for completion times across the five stages of Task 1, comparing the TouchWalker (TW) and virtual joystick (VJ) input conditions. For each stage and input method, we report the mean, median, standard deviation (SD), and interquartile range (IQR).

{\color{rv}

\begin{table}[h]
\centering
\caption{Inferential statistics for completion time comparisons between TouchWalker (TW) and virtual joystick (VJ) for each stage of Task 1.}
\label{tab:completion-time-tests}
\begin{tabular}{lcccc}
\toprule
\textbf{Stage} & \textbf{Test} & \textbf{Test Statistic} & \textbf{p-value} & \textbf{Effect Size} \\
\midrule
Stage 1 & Paired t-test & t = 2.69 & 0.0187* & 0.718 \\
Stage 2 & Wilcoxon & W = 50 & 0.9032 & 0.042 \\
Stage 3 & Paired t-test & t = 3.64 & 0.0030** & 0.973 \\
Stage 4 & Wilcoxon & W = 29 & 0.1531 & 0.394 \\
Stage 5 & Paired t-test & t = 2.90 & 0.0124* & 0.775 \\
\bottomrule
\end{tabular}
\end{table}

Table~\ref{tab:completion-time-tests} summarizes the results of inferential statistical analyses comparing completion times between the TouchWalker (TW) and virtual joystick (VJ). For each stage, we report the statistical test used (paired t-test or Wilcoxon signed-rank test), the corresponding test statistic, p-value, and effect size.
}

\section{Non-Time Performance Metric Statistics}

\begin{table}[h]
\centering
\caption{Non-time performance metric statistics for Task 1 and Task 2 under TouchWalker (TW) and virtual joystick (VJ) input.}
\label{tab:non-time-metric}

\scalebox{0.85}{
\begin{tabular}{llrrrr}
\toprule
\textbf{Task} & \textbf{Method} & \textbf{Mean} & \textbf{Median} & \textbf{SD} & \textbf{IQR} \\
\midrule
\multirow{2}{*}{T1-S1 Falls} & TW & 0.29 & 0.00 & 1.07 & 0.00 \\
 & VJ & 0.07 & 0.00 & 0.27 & 0.00 \\
\multirow{2}{*}{T1-S2 Speed Violations(s)} & TW & 6.90 & 2.54 & 11.40 & 4.17 \\
 & VJ & 12.30 & 2.40 & 21.46 & 11.49 \\
\multirow{2}{*}{T1-S3 Falls} & TW & 2.00 & 1.00 & 2.29 & 3.50 \\
 & VJ & 0.79 & 0.00 & 2.16 & 0.00 \\
\multirow{2}{*}{T1-S5 Collisions} & TW & 6.86 & 7.50 & 3.42 & 3.75 \\
 & VJ & 3.57 & 2.50 & 2.85 & 3.50 \\
\multirow{2}{*}{T2 Water Contacts} & TW & 21.14 & 22.00 & 6.94 & 9.75 \\
 & VJ & 38.29 & 37.00 & 6.31 & 4.50 \\
\bottomrule
\end{tabular}
}

\end{table}

Table~\ref{tab:non-time-metric} summarizes descriptive statistics for the non-time-based performance metrics collected during the user study. These include the number of falls (S1, S3), duration of speed violations (S2), number of collisions with obstacles (S5), and the number of water contacts (T2). Each metric is reported separately for the TouchWalker (TW) and Virtual Joystick (VJ) conditions. Mean, median, standard deviation (SD), and interquartile range (IQR) are provided to characterize overall performance and variability across participants.

{\color{rv}
\begin{table}[h]
\centering
\caption{Inferential statistics for non-time performance metrics comparing TouchWalker (TW) and virtual joystick (VJ) input. For each metric, the statistical test used, test statistic, p-value, and effect size are reported.}
\label{tab:non-time-metric-tests}

\scalebox{0.85}{
\begin{tabular}{lcccc}
\toprule
\textbf{Metric} & \textbf{Test} & \textbf{Test Statistic} & \textbf{p-value} & \textbf{Effect Size} \\
\midrule
T1-S1 Falls & Wilcoxon & W = 1 & 0.6547 & 0.864 \\
T1-S2 Speed Violations(s) & Wilcoxon & W = 27 & 0.1961 & 0.428 \\
T1-S3 Falls & Wilcoxon & W = 11 & 0.0894 & 0.696 \\
T1-S5 Collisions & Wilcoxon & W = 9.5 & 0.0116* & 0.721 \\
T2 Water Contacts & Wilcoxon & W = 0 & 0.0001** & 0.881 \\
\bottomrule
\end{tabular}
}

\end{table}

Table~\ref{tab:non-time-metric-tests} reports the results of inferential statistical analyses for the non-time-based performance metrics. For each metric, we indicate the statistical test applied (Wilcoxon signed-rank test), the corresponding test statistic, p-value, and effect size.

}

\begin{figure}[h]
  \centering

  \includegraphics[trim=125 13 125 50, clip, width=0.11\linewidth]{figs/walk1.png}
  \includegraphics[trim=125 13 125 50, clip, width=0.11\linewidth]{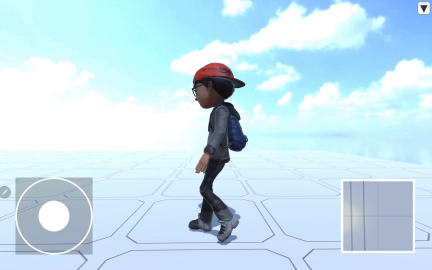} 
  \includegraphics[trim=125 13 125 50, clip, width=0.11\linewidth]{figs/walk3.png}
  \includegraphics[trim=125 13 125 50, clip, width=0.11\linewidth]{figs/walk4.png}
  \includegraphics[trim=125 13 125 50, clip, width=0.11\linewidth]{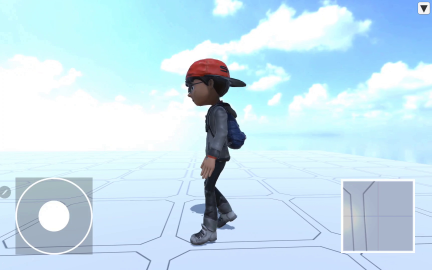} 
  \includegraphics[trim=125 13 125 50, clip, width=0.11\linewidth]{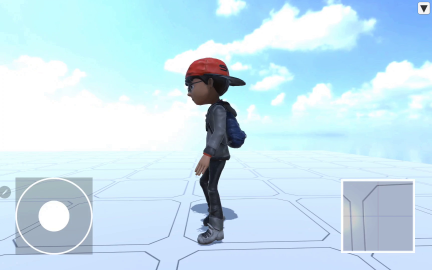} 
  \includegraphics[trim=125 13 125 50, clip, width=0.11\linewidth]{figs/walk7.png}
  \includegraphics[trim=125 13 125 50, clip, width=0.11\linewidth]{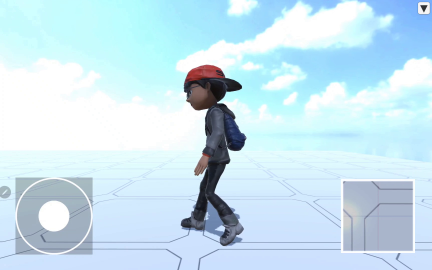}   
\vspace{0pt}

  \includegraphics[trim=100 38 125 63, clip, width=0.11\linewidth]{figs/walk_hand_1.png}
  \includegraphics[trim=100 38 125 63, clip, width=0.11\linewidth]{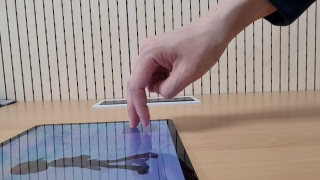} 
  \includegraphics[trim=100 38 125 63, clip, width=0.11\linewidth]{figs/walk_hand_3.png}
  \includegraphics[trim=100 38 125 63, clip, width=0.11\linewidth]{figs/walk_hand_4.png}
  \includegraphics[trim=100 38 125 63, clip, width=0.11\linewidth]{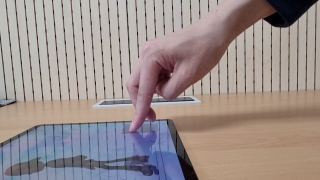} 
  \includegraphics[trim=100 38 125 63, clip, width=0.11\linewidth]{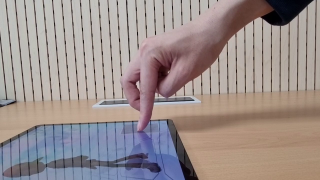} 
  \includegraphics[trim=100 38 125 63, clip, width=0.11\linewidth]{figs/walk_hand_7.png}
  \includegraphics[trim=100 38 125 63, clip, width=0.11\linewidth]{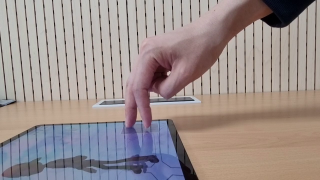}  
\vspace{5pt}
  
  \includegraphics[trim=125 13 125 50, clip, width=0.11\linewidth]{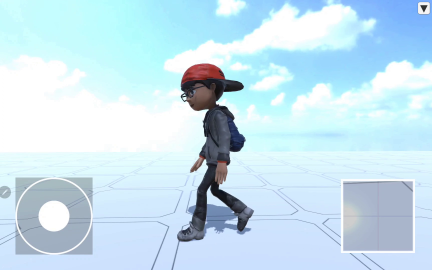}
  \includegraphics[trim=125 13 125 50, clip, width=0.11\linewidth]{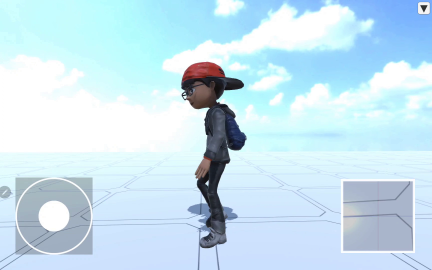} 
  \includegraphics[trim=125 13 125 50, clip, width=0.11\linewidth]{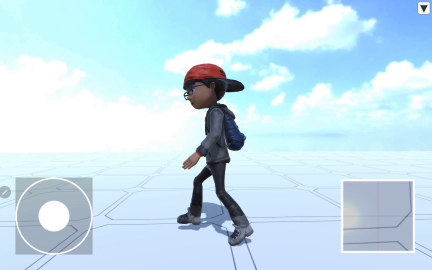}
  \includegraphics[trim=125 13 125 50, clip, width=0.11\linewidth]{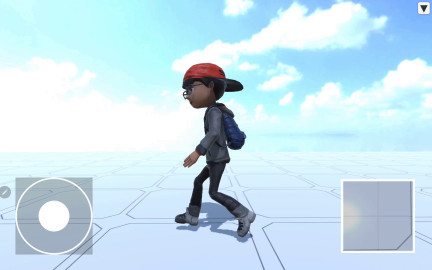}
  \includegraphics[trim=125 13 125 50, clip, width=0.11\linewidth]{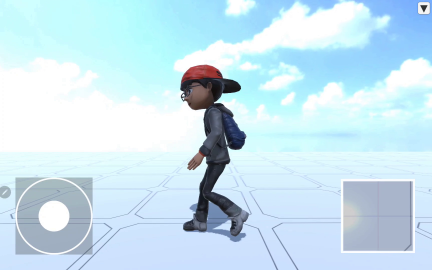} 
  \includegraphics[trim=125 13 125 50, clip, width=0.11\linewidth]{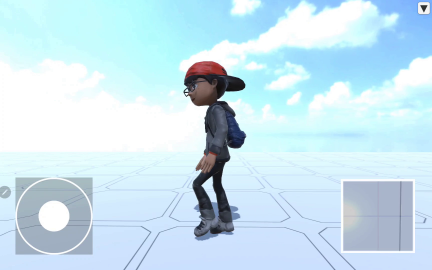} 
  \includegraphics[trim=125 13 125 50, clip, width=0.11\linewidth]{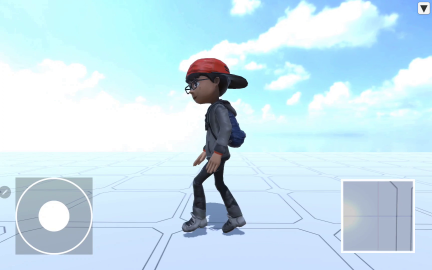}
  \includegraphics[trim=125 13 125 50, clip, width=0.11\linewidth]{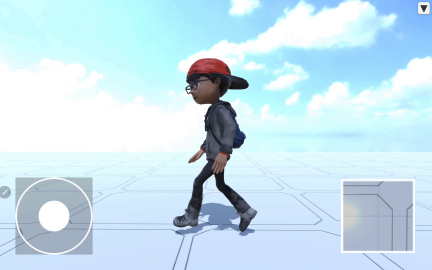} 
\vspace{0pt}

  \includegraphics[trim=100 38 125 63, clip, width=0.11\linewidth]{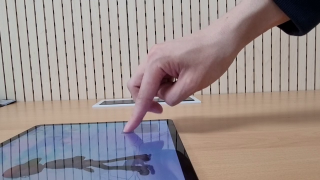}
  \includegraphics[trim=100 38 125 63, clip, width=0.11\linewidth]{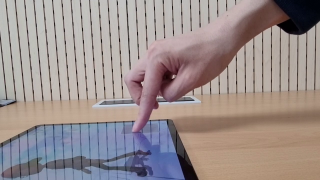} 
  \includegraphics[trim=100 38 125 63, clip, width=0.11\linewidth]{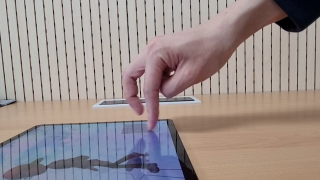}
  \includegraphics[trim=100 38 125 63, clip, width=0.11\linewidth]{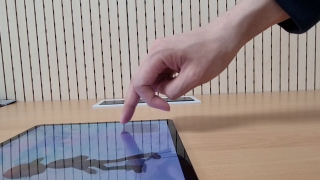}
  \includegraphics[trim=100 38 125 63, clip, width=0.11\linewidth]{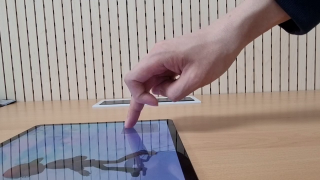} 
  \includegraphics[trim=100 38 125 63, clip, width=0.11\linewidth]{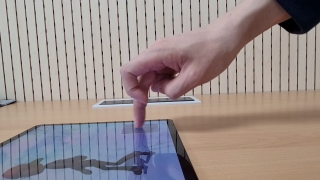} 
  \includegraphics[trim=100 38 125 63, clip, width=0.11\linewidth]{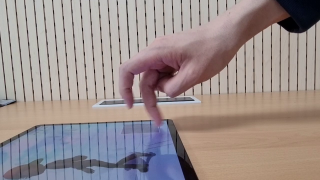}
  \includegraphics[trim=100 38 125 63, clip, width=0.11\linewidth]{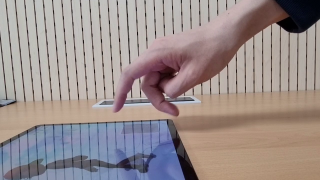}  
\vspace{5pt}
 
  \includegraphics[trim=125 13 125 50, clip, width=0.11\linewidth]{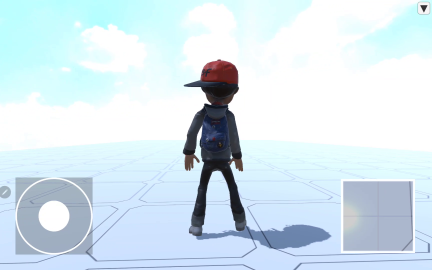}
  \includegraphics[trim=125 13 125 50, clip, width=0.11\linewidth]{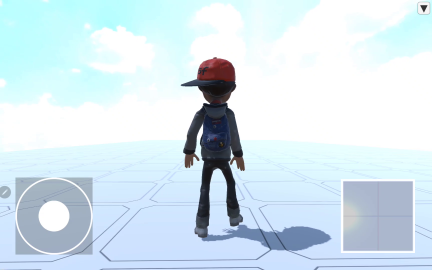} 
  \includegraphics[trim=125 13 125 50, clip, width=0.11\linewidth]{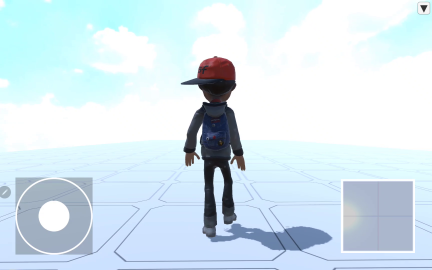}
  \includegraphics[trim=125 13 125 50, clip, width=0.11\linewidth]{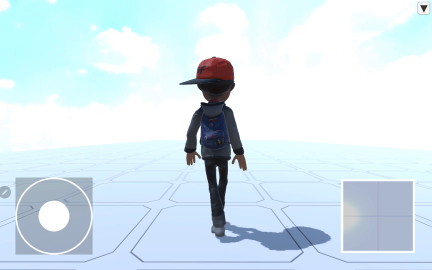}
  \includegraphics[trim=125 13 125 50, clip, width=0.11\linewidth]{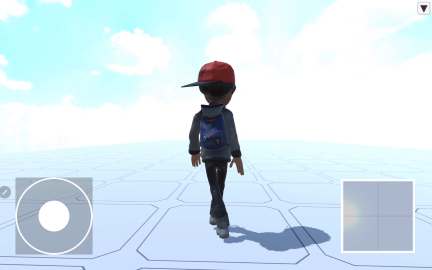} 
  \includegraphics[trim=125 13 125 50, clip, width=0.11\linewidth]{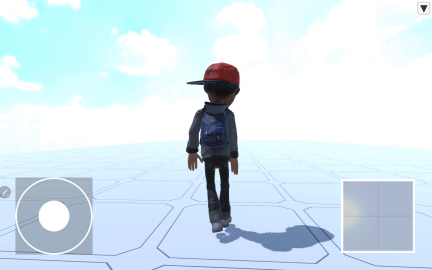} 
  \includegraphics[trim=125 13 125 50, clip, width=0.11\linewidth]{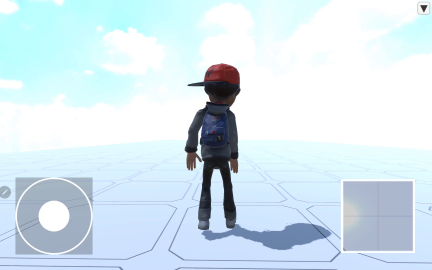}
  \includegraphics[trim=125 13 125 50, clip, width=0.11\linewidth]{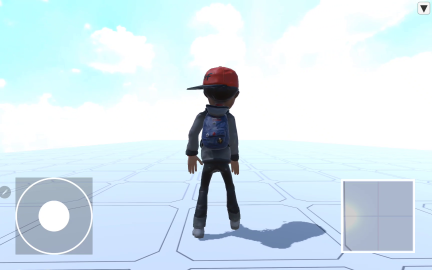}
\vspace{0pt}

  \includegraphics[trim=75 0 100 50, clip, width=0.11\linewidth]{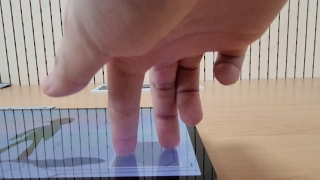}
  \includegraphics[trim=75 0 100 50, clip, width=0.11\linewidth]{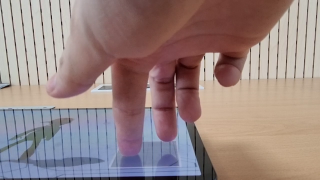} 
  \includegraphics[trim=75 0 100 50, clip, width=0.11\linewidth]{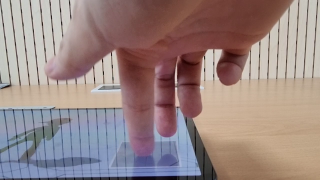}
  \includegraphics[trim=75 0 100 50, clip, width=0.11\linewidth]{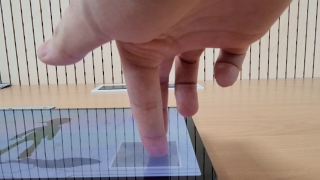}
  \includegraphics[trim=75 0 100 50, clip, width=0.11\linewidth]{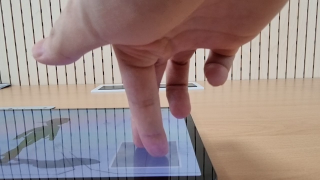} 
  \includegraphics[trim=75 0 100 50, clip, width=0.11\linewidth]{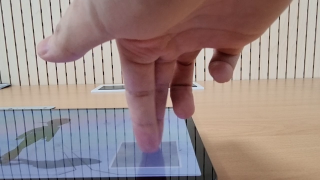} 
  \includegraphics[trim=75 0 100 50, clip, width=0.11\linewidth]{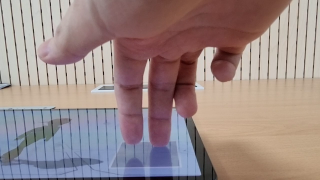}
  \includegraphics[trim=75 0 100 50, clip, width=0.11\linewidth]{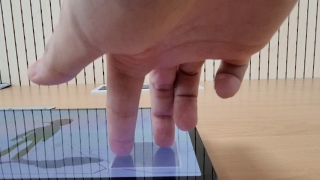} 
\vspace{5pt}

  \caption{Various locomotion styles generated by our system and corresponding finger-walking gesture: walking (row 1), running (row 2), and sideways walking (row 3).
  }
  \label{fig:locomotion}
\end{figure}

{\color{rv2}
\section{Preferred Touch Area Size Analysis}

In the TW condition, participants additionally selected a preferred square touch area size from five candidates ranging from $\times$0.5 to $\times$1.5 relative to a 6\,cm baseline. 
None of the participants selected the smallest or largest sizes (Size~1 or Size~5), resulting in three groups (Sizes~2--4). 
We applied a non-parametric Kruskal--Wallis test to compare both subjective ratings and objective metrics among the three size groups. 

\paragraph{Subjective Ratings}
The results revealed no significant differences in embodiment ($H=0.72, p=0.70$), enjoyment ($H=0.35, p=0.84$), or immersion ($H=2.19, p=0.34$). 
These findings suggest that variations in the selected touch area size did not affect participants’ subjective experience of embodiment, enjoyment, or immersion.

\paragraph{Objective Metrics}
We also compared the objective metrics, including task completion times, number of falls, collision counts, and speed violation durations, across the three selected touch area sizes. 
No significant differences were found in any objective metrics (all $ps > 0.25$), as summarized in Table~\ref{tab:touchsize_objmetrics}. 
These results suggest that the choice of touch area size did not impact task performance.

\begin{table}[h]
  \centering
  \caption{Kruskal--Wallis results for objective metrics across selected touch area sizes (Sizes~2--4).}

\scalebox{0.85}{  
  \begin{tabular}{lccc}
    \hline
    \textbf{Metric} & \textbf{Test} & \textbf{$p$-value} & \textbf{$H$} \\
    \hline
    T1 Completion Time & Kruskal--Wallis & 0.9005 & 0.2095 \\
    T1-S1 Completion Time & Kruskal--Wallis & 0.4325 & 1.6762 \\
    T1-S2 Completion Time & Kruskal--Wallis & 0.9811 & 0.0381 \\
    T1-S3 Completion Time & Kruskal--Wallis & 0.7587 & 0.5524 \\
    T1-S4 Completion Time & Kruskal--Wallis & 0.6271 & 0.9333 \\
    T1-S5 Completion Time & Kruskal--Wallis & 0.4325 & 1.6762 \\
    T2 Completion Time & Kruskal--Wallis & 0.2537 & 2.7429 \\
    T1-S1 Falls & Kruskal--Wallis & 0.5134 & 1.3333 \\
    T1-S2 Speed Violations(s) & Kruskal--Wallis & 0.4876 & 1.4365 \\
    T1-S3 Falls & Kruskal--Wallis & 0.7129 & 0.6768 \\
    T1-S5 Collisions & Kruskal--Wallis & 0.5282 & 1.2765 \\
    T2 Water Contacts & Kruskal--Wallis & 0.6546 & 0.8474 \\
    \hline
  \end{tabular}
}
  
  \label{tab:touchsize_objmetrics}
\end{table}

\paragraph{Hand Length and Touch Area Size}
We further examined the relationship between participants’ hand length and their selected touch area size using Spearman’s rank correlation. 
The results showed a moderate positive trend (\(\rho = 0.4523\)) but did not reach statistical significance (\(p = 0.1045\)). 
This suggests that participants with larger hands tended to choose slightly larger touch areas, but the effect was not statistically reliable.

}

\section{Qualitative Results: Examples of Locomotion Styles}

We captured a sequence of images demonstrating how several avatar locomotion styles can be controlled through touchscreen input (Figure~\ref{fig:locomotion}).

\textit{Walking} (Figure~\ref{fig:locomotion} row 1) can be generated by users alternately touching the finger-walking region with two fingers while progressively moving the touched positions downward, effectively mimicking the actual walking motion.
By ensuring one finger touches slightly before the other lifts off, users can create a double stance phase similar to that seen in real walking.
Users can control the walking characteristics of the character, such as stride and cadence, by adjusting factors like the distance between their fingers and the frequency of touches.

\textit{Running} (Figure~\ref{fig:locomotion} row 2) can be generated similarly to walking by giving touch inputs, but with the touched positions moving downward much faster and by touching one finger only after the other has completely lifted off the screen.
This style of input eliminates the double stance phase, characteristic of running as opposed to walking.

\textit{Sideways walking} (Figure~\ref{fig:locomotion} row 3) can be simulated by adjusting the direction of touch inputs typically used for forward walking. Specifically, this is achieved by shifting the inputs laterally.

\bibliographystyle{abbrv-doi-hyperref}

\bibliography{ref}